\documentstyle[12pt]{article}

\newcommand{\rsection}[1]{\setcounter{equation}{0}\section{#1}}
\renewcommand{\theequation}{\thesection.\arabic{equation}}

\def\d{{\rm d}}
\def\half{{1\over2}}
\def\third{{1\over3}}

\def\sixth{{1\over6}}
\def\lambdam{\hat\lambda_0(2m^2)}
\def\lambdaM{\hat\lambda_0(M^2)}
\def\ifactor#1{\ifnum#1>0
\left\{\lambdam + {#1\over32\pi^2}\right\}
\else\ifnum#1<0
\left\{\lambdam - {\number-#1\over32\pi^2}\right\}
\else
\lambdam
\fi\fi}

\def\bF{{F}}
\def\bU{{U}}

\def\N{\nabla}

\def\intdx{\int\d^d x\/}
\def\Sp(#1,#2){\left<#1,#2\right>}

\newcommand{\be}{\begin{equation}}
\newcommand{\ee}{\end{equation}}
\newcommand{\bea}{\begin{eqnarray}}
\newcommand{\eea}{\end{eqnarray}}

\def\Dh{{D}}

\def\lag{{\cal L}}
\def\ln{{\rm ln\/}}
\def\tr{{\rm tr\/}}
\def\Tr{{\rm Tr\/}}
\def\det{{\rm det\/}}
\def\order{{\cal O}}
\def\leftHright(#1,#2){\left(\stackrel{\rightarrow}{D#1} H_0
            \stackrel{\leftarrow}{D#2}\right)}
\def\leftleftH(#1,#2){\left(\stackrel{\rightarrow}{D#1}
                               \stackrel{\rightarrow}{D#2} H_0\right)}
\def\leftH(#1){\left(\stackrel{\rightarrow}{D#1} H_0\right)}
\def\rightH(#1){\left(H_0\stackrel{\leftarrow}{D#1}\right)}

\newcommand{\artref}[4]{{\sc #1} {\it #2} {\bf #3} #4}

\newcommand{\bookref}[2]{{\sc #1}, #2}

\begin{document}

\begin{titlepage}

\def\mytoday#1{{ } \ifcase\month \or
 January\or February\or March\or April\or May\or June\or
 July\or August\or September\or October\or November\or December\fi
 \space \number\year}

\vspace*{-1.8cm}
\indent\hspace*{10.5cm}\mbox{BUTP-94/12}\newline
\indent\hspace*{10.5cm}\mbox{HUTP-94/A012}
\vspace*{0.3cm}
\begin{center}
{\LARGE Effective Field Theory of the Linear $O(N)$ Sigma Model}
\footnote{Work supported in part by Deutsche Forschungsgemeinschaft,
	by Schweizerischer Nationalfonds,
        and by NSF grant Phy-92-18167.} \\ [1.0cm]

   A. Nyf\/feler

\vspace*{0.1cm}
    Institute for Theoretical Physics \\
    University of Bern \\
    Sidlerstrasse 5, CH-3012 Bern, Switzerland \\[0.4cm]

   and \\[0.4cm]

   A. Schenk

\vspace*{0.1cm}
    The Physics Laboratories\\
    Harvard University \\
    Cambridge, MA 02138, USA \\[0.5cm]

 June 1994 \\
(revised: September 1994)
\vspace*{0.4cm}

\nopagebreak[3]

\begin{abstract}
The low energy structure of a theory containing light and heavy particle
species which are separated by a mass gap can adequately be described by an
effective theory which contains only the light particles. In this work
we present a thorough analysis of the effective field theory of the linear
$O(N)$ sigma model in the spontaneously broken phase.  In particular, we
present a detailed discussion of two techniques, a short-distance expansion and
a method based on loop-integrals,  which can be used to explicitly evaluate the
functional relationships between the low energy constants of the effective
theory and the parameters of the underlying theory. We furthermore provide a
detailed analysis of the matching relation between the linear sigma model
and its effective theory, in order to clarify some discrepancies
which can be found in the literature.
\end{abstract}

\end{center}


\end{titlepage}

\rsection{Introduction}

The effective field theory of the linear sigma model has recently received
considerable attention in the analysis of the symmetry breaking sector of the
electroweak interaction. So far the agreement between experimental data and the
standard model predictions is almost perfect. Yet, the precise nature of the
spontaneous breakdown of the electroweak symmetry remains unknown and various
scenarios are discussed in the literature, including such different theories as
a heavy Higgs~\cite{heavy,Espriu,HRM,EM} or technicolor
models~\cite{techni}. One way out of this ignorance is to replace the Higgs
sector of the standard model by a model independent parametrization which
provides a unified description of the low energy physics in the symmetry
breaking sector, i.e.~up to energies of the order of $1$~TeV. If all the
unknown particles of the underlying theory, e.g.~a Higgs or a Technirho, are
heavy, physics at low energies is dominated by the Goldstone bosons arising
from spontaneous symmetry breaking, and all the other known particles of the
standard model. In this case a convenient description is provided by replacing
the linear sigma model in the standard model Lagrangian by an effective
Lagrangian which contains only the Goldstone
bosons~\cite{heavy,effective2}. The unknown physics
of the symmetry breaking sector is then hidden in the low-energy constants
which occur in the effective Lagrangian. Thus, if the relations between the
constants in the effective Lagrangian and the parameters of the underlying
theory are known,
low energy physics can provide interesting information about
the precise nature of the spontaneous symmetry breaking which may occur at a
comparatively high energy scale, such as in technicolor models.

In this work, we present a thorough analysis of the effective field theory
description of the linear $O(N)$ sigma model in the spontaneously broken phase.
Our goal is twofold. On the one hand, we provide a detailed discussion of the
matching relation between the linear sigma model and its effective theory. A
formal definition of the effective theory as an adequate representation of the
full theory in the low energy region will certainly require that corresponding
Green's functions in both theories have the same low energy structure. It is
known~\cite{matching1,matching2} that literally integrating out only the heavy
degrees of freedom in the full theory generally does not yield an effective
Lagrangian which meets this requirement. Our discussion of this point is
intended to clarify the discrepancies which can be found in the
literature~\cite{GasserLeutChpt,Espriu} on the effective theory of the linear
sigma model. On the other hand, we want to go beyond a formal definition and
provide a detailed discussion of two techniques which can be used to explicitly
evaluate the effective Lagrangian. The first one employs a short distance
expansion in configuration-space, while the second one is based on the
evaluation of loop-integrals in momentum-space.  For the case we are
considering here the result for the effective Lagrangian can already be found
in the literature~\cite{GasserLeutChpt}.  The derivation given there, however,
is rather sketchy and does not discuss the steps involved in any detail. The
present article  contains a thorough account of the problem.  Furthermore it
focuses on the discussion of the  general techniques which can be used to
evaluate the effective Lagrangian for  a given underlying theory, provided the
theory shows a mass gap between heavy  and light particles and the coupling is
weak in the low energy region.

The outline of this work is as follows: In the next section we present a  brief
review of the linear sigma model. In section 3 the general low energy
structure of the effective Lagrangian is discussed, exploiting only the
symmetry  properties of the full theory. In the following section we provide a
thorough analysis of the matching relation between both theories which includes
an exact definition of the effective Lagrangian. In the next three sections
the
detailed description of two methods to calculate the low-energy constants at
order $p^4$ can be found. In section 8 we discuss renormalization and express
the bare quantities in terms of physical parameters. Finally we summarize this
work in section 9.

\rsection{The Linear $O(N)$ Sigma Model}

In this section we employ the technique used in ref.~\cite{GasserLeutChpt}
from which the abundant literature on the linear sigma model may be traced.

In the absence of external fields the Lagrangian of the $O(N)$ symmetric
linear sigma model is given by
\begin{equation}
\half \partial_\mu\phi^T\partial^\mu\phi + \half m^2\phi^T\phi -
{g\over4}\left(\phi^T\phi\right)^2\ ,
\end{equation}
where $\phi^A$ is an $N$-component field. For $m^2>0$ the classical potential
has its minimum at a nonzero value $\phi^T\phi=m^2/g$ and the $O(N)$
symmetry is spontaneously broken to $O(N-1)$. Accordingly, the $N$-component
field $\phi^A$ describes 1 massive (pseudo) scalar and $N-1$ massless Goldstone
bosons.

In order to obtain the generating functional for the Green's functions we
couple the field $\phi^A$ to a set of external fields with spin 0 and spin 1:
\begin{equation} \label{linlag}
\lag_\sigma = \half \nabla_\mu\phi^T\nabla^\mu\phi + \half m^2\phi^T\phi -
{g\over4}\left(\phi^T\phi\right)^2\ + f^T\phi + h \, \tr F_{\mu\nu}F^{\mu\nu} .
\end{equation}
The spin 1 field is defined in terms of the generators $T^\alpha$ of the Lie
algebra $o(N)$ and the $N(N-1)/2$ gauge fields $F_\mu^\alpha$,
\begin{equation}
	F^{AB}_\mu = (T^\alpha)^{AB} F^\alpha_\mu \  ,
\end{equation}
which are coupled covariantly to the scalar field,
\begin{equation} \label{covderiv}
\nabla_\mu\phi^A = \partial_\mu\phi^A + F^{AB}_\mu \phi^B \ .
\end{equation}
The generating functional $W_\sigma\left[F_\mu,f\right]$ is defined by a path
integral
\begin{equation} \label{linfunc}
	e^{i W_\sigma\left[ F_\mu,f\right]}
  		= \int\d\mu[\phi] e^{i\int\d^dx\lag_\sigma} \ .
\end{equation}
Derivatives of this functional with respect to the spin 0 field $f^A$ generate
Green's functions of the scalar fields $\phi^A$ while derivatives with respect
to the spin 1 field $F_\mu^\alpha$ generate Green's functions of the currents,
\begin{equation} \label{conservedcurrents}
	J^\alpha_\mu = (\partial_\mu \phi)^T T^\alpha \phi \ .
\end{equation}

The last term in the Lagrangian is the trace over the square of the field
strength $F_{\mu\nu}^{AB}$ associated with the nonabelian gauge field
$F_\mu^{AB}$,
\begin{equation} \label{Fdef}
	F_{\mu\nu} = \left[ \nabla_\mu ,
		\nabla_\nu \right] \ .
\end{equation}
In the absence of the spin 1 field $F_\mu$ the generating functional
$W_\sigma[0, f]$ can be rendered finite by a proper renormalization of the mass
$m$, the coupling constant $g$, and the scalar fields $\phi^A$ and $f^A$.
Green's functions of the currents, on the other hand, are more singular at
short distances than Green's functions of the fields. The time ordering of
current operators gives rise to ambiguities which do not occur for the fields
themselves. The corresponding Green's functions are unique only up to contact
terms. This ambiguity is reflected by the presence of the last term in the
Lagrangian~(\ref{linlag}), which enters Green's functions of the currents
through contact terms. The corresponding constant $h$ can be renormalized such
that the full generating functional $W_\sigma[F_\mu,f]$ remains finite after
the regulator is removed.

The generating functional defined in eq.~(\ref{linfunc}) has two interesting
properties. First, note that both the generating functional $W_\sigma$ and the
Lagrangian $\lag_\sigma$ are gauge invariant under local $O(N)$ transformations
of the form
\begin{eqnarray}
     \phi & \rightarrow &  V \phi \ , \nonumber\\
     f & \rightarrow & V f \ , \label{gaugetrafo} \\
     F_\mu & \rightarrow & V F_\mu V^T + V\partial_\mu V^T \ .\nonumber
\end{eqnarray}
This statement is equivalent to the Ward identities of the linear sigma model.
Second, this description also allows for the presence of an explicitly symmetry
breaking term of the form $c^T\phi$ which provides the Goldstone bosons with a
mass proportional to $\sqrt{|c|}$. Keeping this constant small enough, one
still
has a mass gap between the Goldstone bosons and the heavy particle. To
obtain Green's functions in this case one merely has to expand the generating
functional around the nonzero value $f=c$ instead of $f=0$.
Except for some minor changes our analysis remains valid for arbitrary values
of the constant $c$, as long as a mass gap is maintained between the Goldstone
bosons and the heavy particle.

Let us pause for a moment and consider the particular case $N=4$ which
describes the symmetry breaking sector in the standard model of the electroweak
interaction. The group $O(4)$ is semi-simple, $O(4)\approx SU(2)\times SU(2)$,
and one can express the nonabelian gauge field $F_\mu$ in terms of external
vector and axial-vector fields:
\begin{eqnarray}
      F^{0i}_\mu & = & a_\mu^i \\
      F^{ij}_\mu & = & - \epsilon^{ijk} v_\mu^k \ .
\end{eqnarray}

Using the method of steepest decent, the one-loop approximation to the
generating functional is given by
\begin{equation} \label{linfunc1}
 W_\sigma\left[F_\mu,f\right] = \int\d^dx
\lag_\sigma(\phi_0,\partial_\mu\phi_0,F_\nu,f) + {i\over2}\ln\det \tilde D\ ,
\end{equation}
where $\tilde D$ is the differential operator
\begin{equation} \label{diffoplin}
(y,\tilde D y) = \int\d^dx y^T\left(\nabla_\mu\nabla^\mu + \tilde\sigma\right)
y
\end{equation}
with
\begin{equation} \label{sigmatilde}
 \tilde\sigma^{AB} = \left(-m^2 + g \phi^T_0\phi_0\right)\delta^{AB} +
2g\phi^A_0\phi^B_0\ .
\end{equation}
The field $\phi^A_0$ is a solution of the classical equations of motion
\begin{equation} \label{eqmolin}
\nabla_\mu\nabla^\mu\phi_0 = f + m^2\phi_0 - g(\phi^T_0\phi_0)\phi_0\ .
\end{equation}

The one-loop approximation given in eq.~(\ref{linfunc1}) represents the first
two terms
in the expansion of the generating functional in powers of $\hbar$. This
corresponds to expansions of the Green's functions in powers of the coupling
constant $g$. In the next step we will analyze the low energy structure of this
approximation in the range of small energies and momenta, i.e., below the
(physical) mass $M$ of the heavy particle. This will in addition introduce an
expansion in powers of momenta $p^2/M^2$. Since we want to describe the low
energy structure of the linear sigma model in terms of an effective Lagrangian
which contains only the Goldstone bosons, the following
parametrization of the classical solution $\phi_0^A$ will turn out to be
useful:
\begin{equation} \label{fielddef}
\phi_0^A \doteq {m\over\sqrt{g}}RU^A \ ,\qquad U^TU = 1 \ .
\end{equation}
The massless modes are now represented by the $N$-component vector $U^A$,
confined to
the sphere $S_{N-1}$ while the massive mode is described by the radial variable
$R$. It is well known~\cite{Coleman}, that the coordinates of the coset space
related
to the spontaneous symmetry breaking pattern, i.e.~the full group factored by
the unbroken subgroup, provide a convenient representation for the Goldstone
boson fields in the effective Lagrangian. In our case the coset space is
$O(N)/O(N-1)$ which is isomorphic to the sphere $S_{N-1}$. Thus, the
definition~(\ref{fielddef}) turns out to be quite obvious.

In terms of the new field the minimum of the classical potential is located at
$R=1$ and the equations of motion~(\ref{eqmolin}) can be rewritten in the form
\begin{eqnarray} \label{eqmolinR}
   \Box R + R(U^T\nabla_\mu\nabla^\mu U) & = & \chi_0^T U + m^2 R ( 1 - R^2) \\
\label{eqmolinU}
    R\left(\nabla_\mu\nabla^\mu U - U (U^T\nabla_\mu\nabla^\mu U)\right)
 & = & \chi_0 - U(\chi_0^TU) - 2 \partial_\mu R\nabla^\mu U \ ,
\end{eqnarray}
where
\begin{equation} \label{chidef}
\chi_0\doteq {\sqrt{g}\over m} f \ .
\end{equation}

The radial variable $R$ describes a massive fluctuation around $R=1$ and its
value at a particular space-time point depends only on the behaviour of the
external field in a small neighbourhood of this point. Thus,
for slowly varying external fields the behaviour of the massive mode is under
control and the equation of motion~(\ref{eqmolinR}) can be solved
algebraically. The
result is an expansion in powers of derivatives of the external fields:
\begin{equation} \label{rexp}
R = 1 + \delta R_1 + \delta R_2 + \ldots \ ,\qquad \delta R_n = \order(p^{2n})
\ .
\end{equation}
Note that the field $f$ counts as a quantity of order $p^2$ while the gauge
field $F_\mu$ occurring in the covariant derivative is of order $p$. $U$ is a
quantity of order~1.

The first two nontrivial terms of this series are readily evaluated to be
\begin{eqnarray}
     \delta R_1 & = & {1\over2m^2}\left[\chi^T_0U + \nabla_\mu U^T\nabla^\mu
U\right] \label{rone} \\
      \delta R_2 & = & -{3\over 2} \delta R_1^2 + {1\over 2 m^2} \delta R_1
(\nabla_\mu U^T\nabla^\mu U) + \mbox{\rm total derivative} \ .\label{rtwo}
\end{eqnarray}
The tree-level contribution to the generating functional is given by
\begin{equation}
   \half{m^2\over g}\int\d^dx \left(R (\chi_0^TU)+\half m^2 R^4\right) + h
\int\d^dx\tr F_{\mu\nu} F^{\mu\nu} \ .
\end{equation}
Thus, with the help of expansion~(\ref{rexp}) one is able to determine the low
energy behaviour of the tree-level contribution to the generating functional to
any required order in $p^2$.

Finally, by changing to a new basis, we separate the massless fluctuations
around the classical solution $\phi^A_0$, which are bound to the sphere
$U^TU=1$, from the massive fluctuation along $U^A$:
\begin{equation}
  \phi^A-\phi^A_0 = \xi U^A + \sum_{i=1}^{N-1}\epsilon_i^A\eta^i \ ,
\end{equation}
with
\begin{equation} \label{ortho}
   \epsilon_i^T\epsilon_j  = \delta_{ij} \ ,\qquad \epsilon_i^T U = 0 \ ,
\end{equation}
and with the completeness relation
\begin{equation} \label{vollst}
   U^A U^B + \sum_{i=1}^{N-1}\epsilon^A_i\epsilon^B_i = \delta^{AB} \ .
\end{equation}
In order to describe the action of the differential operator $\tilde D$ in the
new basis of the mass eigenstates we introduce the following notation:
\begin{eqnarray} \label{linopsdef1}
d          	&=&  \Box + 2 m^2 - f_\mu^T f^\mu  + 3 m^2 (R^2 - 1)\\
f_i^\mu         &=&   U^T\nabla^\mu\epsilon_i \\
D_\mu 		&=&     \partial_\mu + \Gamma_\mu \\
\Gamma_\mu^{ij}	&=&	\epsilon_i^T\nabla_\mu\epsilon_j\\
\delta  	&=&	(D_\mu f^\mu) + 2 f_\mu^T D^\mu \label{def_delta} \\
(\eta,\delta^T \xi) &= & (\delta \eta,\xi) \label{def_deltaT} \\
D 		&=&	D_\mu D^\mu  + \sigma \label{def_D} \\
\sigma^{ij}     &=&    - f_\mu^i f^{j \mu} + m^2(R^2-1)\delta^{ij} \ .
\label{linopsdef2}
\end{eqnarray}
One finds
\begin{equation}
(y,\tilde D y) = (\xi,d \xi) + (\xi,\delta\eta) + (\delta^T\xi,\eta) + (\eta,
D\eta) \ .
\end{equation}
To separate massless and massive fluctuations we furthermore diagonalize
this expression with the transformation $\hat\xi \doteq \xi +
d^{-1}\delta\eta$:
\begin{equation}
(y,\tilde D y) = (\hat\xi,d\hat\xi) + \left(\eta, (D -
\delta^Td^{-1}\delta)\eta\right) \ .
\end{equation}
Thus, we obtain the following representation for the one-loop contribution to
the generating functional
\begin{equation} \label{tildeexpansion}
     \ln\det \tilde D  = \ln\det D + \ln\det d
                      + \ln\det (1- D^{-1}\delta^Td^{-1}\delta) \ .
\end{equation}

\rsection{The Effective Lagrangian}

In this section we will analyze the general low energy structure of the linear
$O(N)$ sigma model. We will set up the effective Lagrangian
description for the generating functional exploiting only the symmetry
properties of the linear sigma model. Thus, this discussion is completely
general and applies to any underlying theory which has the same symmetry
breaking pattern and a mass gap between the Goldstone bosons and the
heavier particle species. To begin the discussion, let us recall that the
generating functional $W_\sigma[F_\mu,f]$ as defined in the previous section is
gauge invariant under local $O(N)$ transformations as given in
eq.~(\ref{gaugetrafo}). Due to the presence of
Goldstone bosons generated by spontaneous symmetry breaking, the
structure of the Green's functions is nontrivial even at low energies. Only the
contributions from the massive modes admit a Taylor series expansion. It is
known~\cite{Leutwyler} that, if the underlying theory (the linear sigma model
in our
case) does not contain anomalies and is Lorentz invariant, the low energy
structure of the generating functional can be described in terms of an
effective field theory with a symmetric effective Lagrangian.

In our case, this effective Lagrangian depends on an $N$-component field
$U^A$   confined to the sphere $U^T U=1$ which describes
the $N-1$ Goldstone bosons and which is coupled to the external fields $f^A$
and $F_\mu^{AB}$. The latter is coupled to the $O(N)$ vector $U^A$ by
the covariant derivative given in eq.~(\ref{covderiv}). The effective
Lagrangian is the most general functional invariant under Lorentz
transformations and the gauge transformations given in eq.~(\ref{gaugetrafo}),
\begin{equation}
\lag_{eff} = \lag_{eff}\left(U,\nabla_\mu U,
\nabla_\mu\nabla_\nu U, f, \ldots\right) \ .
\end{equation}
The Lagrangian $\lag_{eff}$ is a sum of terms with an increasing
number of derivatives, corresponding to an expansion in powers of the momentum,
\begin{equation}
 \lag_{eff} = \lag_2 + \lag_4 + \lag_6 + \ldots
\end{equation}
where $\lag_i$ is of order $p^{i}$. (If explicitly symmetry breaking terms
are present, this expansion includes powers of the Goldstone boson masses as
well.) The generating functional is defined as the path integral
\begin{equation}
e^{i W_{eff}\left[ F_\mu,f\right]} = \int\d\mu[U] e^{i\int\d^dx\lag_{eff}} \ .
\end{equation}

In general, there are two different kinds of contributions to the generating
functional. On the one hand, we have tree-level contributions, given by the
integral $\int\d^dx\lag_{eff}$, which has to be evaluated at the stationary
point, i.e.~with the solutions of the equations of motion. One the other hand
there are contributions from loops, which ensure unitarity. General power
counting arguments show~\cite{Weinberg}, that $n$-loop corrections are
suppressed by
powers $p^{2n}$ as compared to the tree-level. Thus, one-loop corrections with
vertices of $\lag_2$ are of order $p^4$ while those with vertices of $\lag_4$
are of order $p^6$, as are two-loop corrections. The suppression of loops by
powers of $p^2$ which allows the perturbative study of the systematic expansion
in powers of momenta
is related to the fact that Goldstone bosons do not interact at
threshold. Thus, at order $p^4$ the generating functional $W_{eff}[F_\mu,f]$ is
given by
\begin{equation} \label{genfuncp4}
     W_{eff}\left[F_\mu,f\right] = \int\d^dx\>\left(\lag_2 + \lag_4\right) +
{i\over2}\ln\det \bar D + \order(p^6) \ ,
\end{equation}
where the term $\ln\det \bar D$ describes the one-loop corrections with
vertices from $\lag_2$.

The leading term of the generating functional is of order $p^2$ and given by
the tree-level contributions of $\lag_2$
\begin{equation} \label{genfuncp2}
W_{eff}\left[F_\mu,f\right] = \int\d^dx \lag_2 + \order(p^4) \ .
\end{equation}
Imposing Lorentz invariance and taking account of the identities $U^T
\nabla_\mu U = 0$ and $U^T\nabla_\mu\nabla^\mu U = - \nabla_\mu U^T
\nabla^\mu U$, the most general effective Lagrangian at order $p^2$ which
is invariant under local $O(N)$ transformations
involves only two low energy constants and is given by
\begin{equation}
   c_1 \nabla_\mu U^T\nabla^\mu  U + c_2 f^T U \ .
\end{equation}
In view of eq.~(\ref{chidef})
it will turn out to be convenient to introduce another field $\chi^A$
proportional to $f^A$ and define this Lagrangian in the form
\begin{equation} \label{eqmonlag}
\lag_2 = F^2\left(\half\nabla_\mu U^T\nabla^\mu U + \chi^T U\right) \ .
\end{equation}
Accordingly, we will replace the dependence of the generating functional
$W_{eff}$ and of the effective Lagrangian $\lag_{eff}$ on the field $f^A$ by
the dependence on the field $\chi^A$.
The action in eq.~(\ref{genfuncp2}) is to be evaluated with the solution
$\bar U^A$ of the classical equations of motion
\begin{equation} \label{eqmonlin}
    \nabla_\mu\nabla^\mu \bar U - \bar U (\bar U^T\nabla_\mu\nabla^\mu
\bar U) =  \chi - \bar U (\chi^T\bar U) \ .
\end{equation}
Note that this solution coincides with the solution $U^A$ of
eq.~(\ref{eqmolinU}) only
at leading order. The corrections $U^A - \bar U^A$ are due to the massive
mode and admit an expansion in powers of $p^2$ (see below).

To obtain the one-loop contribution of order $p^4$ we again use the method of
steepest descent and parametrize the fluctuations around the solution $\bar
U^A$ of the equations of motion in a way similar to the one used in the
previous section on the linear sigma model,
\begin{equation} \label{xxx}
          U^A - \bar U^A = \xi^i \bar\epsilon_i^A - \half \xi^i\xi^i\bar
U^A + \ldots \ .
\end{equation}
The vectors $\{\bar U^A,\bar\epsilon^A_i\}$ are orthonormal, and thus obey
relations as given in eqs.~(\ref{ortho},\ref{vollst}). The presence of the
second term in eq.~(\ref{xxx}) is necessary to ensure that both vectors, $\bar
U^A$ and $U^A$, are of unit length. Introducing a notation similar to the
definitions~(\ref{linopsdef1}-\ref{linopsdef2}),
\begin{eqnarray}
     \bar f^\mu_i &=& \bar U^T\nabla^\mu \bar\epsilon_i\\
     \bar D_\mu & = & \partial_\mu + \bar\Gamma_\mu \\
 \bar\Gamma_\mu^{ij} & = & \bar\epsilon^T_i\nabla_\mu\bar\epsilon_j \\
     \bar\sigma^{ij} & = &  - f^i_\mu f^{j \mu} +
\left(\nabla_\mu\bar U^T\nabla^\mu\bar U + \chi^T\bar U\right) \delta^{ij} \ ,
\end{eqnarray}
the differential operator $\bar D$ in eq.~(\ref{genfuncp4}) turns out to be
\begin{equation}
   \bar D = \bar D_\mu \bar D^\mu + \bar \sigma \ .
\end{equation}

Finally, we have to specify the general effective Lagrangian at order $p^4$.
Note that the most general effective Lagrangian at this order is given as a
linear combination of a maximal set of independent $O(N)$ invariant terms of
order $p^4$. On the one hand, redundant terms can be eliminated by using
relations of the form
\begin{equation}
\int \d^dx \N_\nu\N_\mu\bU^T\N^\mu\N^\nu\bU
        = \int \d^dx \N_\mu\N^\mu\bU^T\N_\nu\N^\nu\bU +
                                \N_\nu\bU^T\bF^{\nu\mu}\N_\mu\bU \ ,
\end{equation}
which are readily verified using the definition of the field strength given in
eq.~(\ref{Fdef}). On the other hand, the Lagrangian $\lag_4$ contributes only
at the classical level. Hence, the equations of motion~(\ref{eqmonlin}) can
also  be  used to eliminate further redundant terms. From eq.~(\ref{eqmonlin})
one can infer the  following two identities
\begin{eqnarray} \label{eqmocond}
 (\chi^T\nabla_\mu\nabla^\mu\bar U) +  (\chi^T\bar U)(\N_\mu\bar U^T\N^\mu\bar
U)
 + (\chi^T \bar U)^2  & = & \chi^T \chi \\
 (\nabla_\mu\nabla^\mu\bar U^T \nabla_\nu \nabla^\nu\bar U) -
 (\N_\mu\bar U^T\N^\mu\bar U)^2  + (\chi^T \bar U)^2  & = & \chi^T \chi \ .
 \label{eqmocond2}
\end{eqnarray}
The general effective Lagrangian at order $p^4$ contains eight independent low
energy constants and can be brought into the form
\begin{eqnarray} \label{nonlag4}
\lefteqn{\lag_4 = d_1 (\N_\mu\bar U^T\N^\mu\bar U)^2 +
d_2 (\N_\mu\bar U^T\N_\nu\bar U)(\N^\mu\bar U^T\N^\nu\bar U)} \nonumber \\
& &\qquad\mbox{}
+ d_3 (\chi^T\bar U)^2 + d_4(\chi^T\bar U)(\N_\mu\bar U^T\N^\mu\bar U)
\nonumber \\
& & \qquad\mbox{}  + d_5 (\bar U^T F_{\mu\nu}F^{\mu\nu}\bar U) +
d_6 (\N_\nu\bar U^T F^{\nu\mu}\N_\mu\bar U) \nonumber \\
& & \qquad\mbox{} + d_7 (\chi^T \chi) + d_8 \tr(F_{\mu\nu} F^{\mu\nu}) \ .
\end{eqnarray}

\rsection{Matching}

In section two we have determined the generating functional of the linear sigma
model up to one loop. In the next section we have described the low energy
structure of this generating functional in terms of an effective Lagrangian
which contains only the $N-1$ Goldstone bosons. Exploiting the
symmetry properties of the underlying theory one can reduce the number of low
energy constants that are involved in the effective Lagrangian to two at order
$p^2$ and to eight at order $p^4$. All these low energy constants are
determined by the underlying theory. To express the low energy constants in
terms of the parameters of the linear sigma model we require that both theories
yield the same Green's functions in the low energy region:
\begin{equation}\label{Zequal}
W_\sigma\left[F_\mu,f\right] = W_{eff}\left[F_\mu,\chi\right] \ .
\end{equation}
Recall that we have absorbed one of the low energy constants in section~3 by
introducing the external scalar field $\chi^A$. Thus eq.~(\ref{Zequal})
determines the relation between the scalar fields $\chi^A$ and $f^A$ as well.
Using eqs.~(\ref{linfunc1},\ref{tildeexpansion}) and (\ref{genfuncp4}) we
obtain at order $p^4$:
\begin{eqnarray} \label{match1}
\lefteqn{ \int\d^dx\>\lag_\sigma + {i\over2}\ln\det D  +
{i\over2} \ln\det d  + {i\over2} \ln\det (1-
D^{-1}\delta^Td^{-1}\delta)} \hspace{2cm} \nonumber\\
\qquad\qquad& = &  \int\d^dx\>\left(\lag_2 + \lag_4\right) +
{i\over2}\ln\det \bar D \ . \qquad
\end{eqnarray}

We will solve this equation for the low energy constants in a couple of steps.
To begin with, let us recall that both theories, the linear sigma model and the
effective theory, contain Goldstone bosons. Thus, both sides of
eq.~(\ref{match1}) should reproduce all singularities associated with the light
particles and the corresponding contributions should cancel. This is indeed the
case. At order $p^4$ all of these singularities are described by the
determinants of the two differential operators $ D$ and $\bar D$. A
glance at eqs.~(\ref{eqmolinU},\ref{eqmonlin}) shows that there is only a
small difference between
these operators, related to the fact that the fields $U^A$ and $\bar U^A$
satisfy slightly different equations of motion. We will,
however, see below, that the corresponding difference between $\ln\det  D$
and $\ln\det\bar D$ is of higher order and can be neglected in our analysis.
Thus, at order $p^4$ our matching relation simplifies to
\begin{equation} \label{match2}
 \int\d^dx\>\lag_\sigma +
{i\over2} \ln\det d  + {i\over2} \ln\det (1- D^{-1}\delta^Td^{-1}\delta)
=   \int\d^dx\>\left(\lag_2 + \lag_4\right) \ .
\end{equation}
Note that the two determinants on the left hand side of this equation are
still nonlocal objects, while the right hand side is a local expression.
The
operator $d$, however, is related to the heavy particle in the linear sigma
model and its determinant admits an expansion in terms of local quantities
involving an increasing number of derivatives, which correspond to an expansion
in powers of momenta. The second determinant on the left hand side of
eq.~(\ref{match2}) is a more complicated matter. Since it involves the operator
$D$ of the massless modes it is not a purely local object. The analysis in the
following reveals that its nonlocality in the low-energy region shows up at
the order $p^6$. Thus, at the order $p^4$ all terms in the low energy expansion
of the determinants in eq.~(\ref{match2}) are local. The comparison of
corresponding coefficients on both sides of this equation yields the functional
relationship between the low energy constants and the parameters of the linear
sigma model at this order.

If one were to determine the low energy constants up to the order $p^6$, the
generating functionals of the linear sigma model and of the effective theory
would have to be calculated up to the two-loop level. In this case, additional
terms would occur which ensure that the matching condition contains only local
terms up to the order $p^6$. This will be explained in greater detail below.

Before we use eq.~(\ref{match2}) to determine the low-energy constants up to
the order $p^4$, some further comments on our matching relation~(\ref{Zequal})
are in order. The crucial point to notice here is that the expansion of the
nonlocal terms in eq.~(\ref{match2}) takes place after all integrations over
the fields $U$ and $R$ have been performed. More explicitly, this amounts to a
relation between the full path integrals, i.e.
\begin{equation} \label{fullmatch}
\int\d\mu[U] e^{i\int\d^dx\lag_{eff}(U,F_\mu,\chi)}
              = \int\d\mu[\phi] e^{i\int\d^dx\lag_\sigma(\phi,F_\mu,f)} \ .
\end{equation}
where $\lag_{eff} = \lag_2 + \lag_4$ at order $p^4$. To get a
better
understanding of this point, let us define the quantity
$\Gamma_{eff}[U,F_\mu,f]$ (as in ref.~\cite{Espriu}) by performing the path
integral on the right hand
side of eq.~(\ref{fullmatch}) only over the massive particle,
\begin{equation} \label{def_Gamma}
e^{i \Gamma_{eff}\left[U,F_\mu,f\right]} \doteq \int\d\mu[R]
e^{i\int\d^dx\lag_\sigma(R,U,F_\mu,f)} \ ,
\end{equation}
where we have used the parametrization~(\ref{fielddef}) for the field $\phi^A$
as well as the identity
\begin{equation}
  \d\mu[\phi] = \d\mu[U]\d\mu[R] \ .
\end{equation}
Our matching relation can thus be rewritten in the form
\begin{equation} \label{fullmatch2nd}
\int\d\mu[U] e^{i\int\d^dx\lag_{eff}(U,F_\mu,\chi)}
              = \int\d\mu[U] e^{i\Gamma_{eff}\left[U,F_\mu,f\right]} \ .
\end{equation}
Note that the quantity $\Gamma_{eff}[U,F_\mu,f]$ itself is a nonlocal object.
Since it
involves only the propagation of the heavy particle, it certainly admits an
expansion in terms of local quantities as well. Such an expansion would
introduce another density $L_{eff}$, defined by
\begin{equation}
\int\d^dx L_{eff} = \Gamma_{eff} \ .
\end{equation}
However, in eq.~(\ref{fullmatch2nd}), which defines the effective Lagrangian
$\lag_{eff}$, one first integrates over the massless modes
$U$ and then expands the resulting nonlocal contributions, as explained
in the sequel of eq.~(\ref{match2}). One may now pose the question of whether
it
is permissible to change the order of these two operations. In that case one
would first expand the integrand on the right hand side of
eq.~(\ref{fullmatch2nd}), and then perform the integration over the massless
modes. If that procedure were correct, one
would obtain the result that $L_{eff}$ and $\lag_{eff}$ are the same and
our matching relation~(\ref{fullmatch2nd}) between integrals could be
replaced be an equivalent relation between integrands, i.e.
\begin{equation}
   e^{i\int\d^dx \lag_{eff}(U,F_\mu,\chi)}  = \int\d\mu[R]
e^{i\int\d^dx\lag_\sigma(R,U,F_\mu,f)} \ .
\end{equation}
At the classical level this relation is obviously correct, since then
\begin{equation}
\int\d^dx\lag_{eff} = \int\d^dx L_{eff} = \Gamma_{eff} = \int\d^dx\lag_\sigma
\ ,
\end{equation}
where $\lag_\sigma$ is to be evaluated with the classical solution $R$ of the
equations of motion~(\ref{eqmolinR}). However, if quantum corrections are taken
into account, it is generally wrong.
Using the method of steepest decent, we again expand the field $R$ as
\begin{equation}
	R = R_0 + z \ ,
\end{equation}
where $R_0$ satisfies the equations of motion~(\ref{eqmolinR}). The measure is
given by
\begin{equation}
      	\d\mu[R] = {\cal N} \prod_x R^{N-1} \d R = {\cal N} \prod_x R_0^{N-1}
	(1+\order(z)) \d z \ .
\end{equation}
Furthermore, in the dimensional regularization scheme the factor $R_0^{N-1}$
in the measure does not contribute to the integral. Thus, at the order we are
considering here, the measure is given by
\begin{equation}
	\d\mu[R] = {\cal N} \prod_x \d z \ ,
\end{equation}
and one obtains
\begin{equation} \label{wrongmatch}
 \int\d^dx\>\lag_\sigma +
{i\over2} \ln\det d  =   \int\d^dx\> L_{eff} \ .
\end{equation}
Now one can compare the results for $L_{eff}$ and $\lag_2 + \lag_4$ as given in
eqs.~(\ref{wrongmatch}) and~(\ref{match2}). At order $p^4$ they both receive
contributions from loop integrals which include only propagators of massive
particles, represented by the term $\ln\det d$. In addition to that, however,
$\lag_2+\lag_4$ furthermore receives contributions from loop integrals which
include propagators of both, massive and massless particles, as described by
the
last term on the left hand side of eq.~(\ref{match2}). These contributions are
missing in the representation~(\ref{wrongmatch}) of $L_{eff}$. Thus, if quantum
corrections are taken into account, one must require equality between the full
integrals, as in our matching condition~(\ref{Zequal}). In general this
requirement cannot be replaced by a relation between integrands. In other
words, the integration over the massless modes and the expansion of the
nonlocal objects do not commute. The relevance of mixed loops containing
propagators of light and heavy particles is also discussed in the framework
of QED in ref.~\cite{matching2}.

\rsection{Short-Distance Expansion}

In this section we will discuss the evaluation of the low-energy constants from
eq.~(\ref{match2}) with the help of short-distance expansions.
The first term on the left hand side of eq.~(\ref{match2}), which describes all
tree-level contributions, is readily evaluated with the help of
eqs.~(\ref{rexp},\ref{rone},\ref{rtwo})
\begin{eqnarray} \label{linclass}
\lefteqn{\int\d^d x \lag_\sigma = h \int\d^dx \tr (F_{\mu\nu}F^{\mu\nu})}
\nonumber \\
&+& \left({m^2\over g}\right) \int\d^dx \left[ \half \N_\mu U^T\N^\mu U
+ \chi_0^T U\right] \\
&+& \left({1\over 4g}\right)\int\d^d x \left[ (\N_\mu U^T \N^\mu U)^2
+ 2(\N_\mu U^T \N^\mu U) (\chi_0^T U) + (\chi_0^T U)^2\right] \ . \nonumber
\end{eqnarray}

The second term on the left hand side of eq.~(\ref{match2}) describes all those
one-loop corrections which involve only heavy particles. We expand
this contribution in powers of the difference $\sigma_M = d - (\Box + 2 m^2)$
which counts as a quantity of order $p^2$. At order $p^4$ we get
\begin{equation} \label{lndetd}
 {i\over2} \ln\det d = {i\over2}\ln\det d_0  + {i\over2}\Tr [d_0^{-1}
\sigma_M]
			- {i\over4}\Tr [ d_0^{-1} \sigma_M]^2  \ ,
\end{equation}
with $d_0 = \Box + 2 m^2$. The second term is given by
\begin{equation} \label{lndetd_2}
          {i\over2} G_M(0) \int\d^d x \sigma_M(x) \ ,
\end{equation}
where $G_M(z)$ is the Feynman propagator of a scalar particle with mass
$M=\sqrt{2}m$. The third term is of the form
\begin{equation} \label{lndetd_3}
- {i\over4} \int\d^d x \d^d y \sigma_M(x) \sigma_M(x+y) G_M^2(y)
 =  -{i\over4} \int\d^d x \sigma_M^2(x) \int\d^d y G^2_M(y) + \order(p^6)  \ .
\end{equation}
Note that only the local behaviour of the slowly varying external fields is
relevant since the massive propagator cuts off large distances.
Using eqs.~(\ref{rone},\ref{rtwo},\ref{linopsdef1}) we obtain the following
relevant contribution
\begin{eqnarray}
{i\over 2} \ln\det d =
&-& 2m^2\lambdam\int\d^d x\ \left[ 2 (\N_\mu U^T \N^\mu U) + 3  (\chi_0^T U)
\right] \nonumber \\
&-& \left(2\lambdam+{1\over16\pi^2}\right)
                \int\d^d x \ (\N_\mu U^T \N^\mu U)^2 \nonumber \\
&-& \left(3\lambdam+{3\over16\pi^2}\right)
                \int\d^d x \  (\N_\mu U^T \N^\mu U)(\chi_0^T U) \nonumber \\
&-& \left({3\over2}\lambdam+{9\over64\pi^2}\right)
                 \int\d^dx (\chi_0^T U)^2 \ ,   \label{Delta_1}
\end{eqnarray}
where
\begin{eqnarray}
\hat\lambda_0(M^2) &\doteq& \lambda_0 + {1\over32\pi^2}\ln{M^2\over\mu^2}\\
\lambda_0	&\doteq&  {1\over 16 \pi^2} \mu^{d-4} \left(
       {1\over d-4} - \half \left( \ln 4 \pi + \Gamma' (1) + 1 \right) \right)
\ .
\end{eqnarray}

This leaves us with the last term on the left hand side of eq.~(\ref{match2})
which is
the hardest to evaluate. If we again expand the localized quantity
$\langle x|d^{-1}|y\rangle$ in powers of $\sigma_M = d - (\Box + 2 m^2)$,
only the first three terms contribute at order $p^4$
\begin{eqnarray} \label{Delta_2a}
\lefteqn{ {i\over2}\ln\det (1 -  D^{-1} \delta^T d^{-1} \delta) =
- {i\over2}\Tr [ \delta D^{-1} \delta^T d_0^{-1} ]} \nonumber \\
&& + {i\over2}\Tr [ \delta D^{-1}\delta^T d_0^{-1}\sigma_M d_0^{-1}]
- {i\over4} \Tr [ \delta D^{-1}\delta^T d_0^{-1}]^2 \ ,
\end{eqnarray}
where $d_0 = \Box + 2m^2$.
Furthermore, for the last two terms only the leading singularity of the
kernel $\langle x|\delta D^{-1}\delta^T|y\rangle$, determined by the
d'Alembertian, is relevant at this
order. In particular, the relevant contribution to the second term is given by
\begin{eqnarray} \label{Delta_2a_2}
\lefteqn{ -2i\int\d^dx\d^dy\d^dz f_\mu^T(x) f_\nu(x+y) \sigma_M(x+z)
    \partial^\mu\partial^\nu G_0(y) G_M(y-z) G_M(z) } \nonumber \\
&&\hspace{-0.7cm}=  -2i\int\d^dx f_\mu^T(x) f_\nu(x) \sigma_M(x)
\int\d^dy \d^dz
\partial^\mu\partial^\nu G_0(y) G_M(y-z) G_M(z) + \order(p^6),
\nonumber \\[-0.2cm]
&&
\end{eqnarray}
where $G_0(z)$ is the Feynman propagator of a massless particle.
Again the massive propagator cuts off large distances.
Lorentz invariance requires this contribution to be of the form
\begin{equation}
-{2 i \over d} \int \d^dx f_\mu^T f^\mu \sigma_M \int \d^dy G_M^2(y) \ .
\end{equation}
In the same way, we obtain the following result for the relevant contribution
to the third term in eq.~(\ref{Delta_2a})
\begin{eqnarray} \label{Delta_2a_3}
  \lefteqn{\hspace{-0.7cm}-4i \int\d^dx (f_\mu^T f_\nu) (f_\rho^T f_\sigma)
              \int\d^dy \d^dz\d^dv \partial^\mu\partial^\nu G_0(y) G_M(y-z)
			\partial^\rho\partial^\sigma G_0(z-v) G_M(v) }
\nonumber \\
& = & -{4i\over d(d+2)} \int\d^dx\left( (f_\mu^T f^\mu) (f_\nu^T f^\nu) +
               2 (f_\mu^T f^\nu) (f_\nu^T f^\mu)\right) \int\d^dy G_M^2(y) \ .
\nonumber \\[-0.2cm]
&&
\end{eqnarray}
Using the identity
\begin{equation} \label{ffidentity}
         f^T_\mu f_\nu = \nabla_\mu U^T \nabla_\nu U \ ,
\end{equation}
we get at order $p^4$
\begin{eqnarray}\label{d_Delta_2}
\lefteqn{ {i\over2}\Tr [  \delta D^{-1}\delta^T d_0^{-1}\sigma_M d_0^{-1}]
- {i\over4} \Tr [ \delta  D^{-1}\delta^T d_0^{-1}]^2 =} \nonumber \\
&- & \left( {7\over3} \lambdam + {1\over(4\pi)^2} {19\over36} \right)
     \intdx \ (\nabla^\mu U^T\nabla_\mu U)^2 \nonumber\\
&-& \left( {2\over3} \lambdam + {1\over(4\pi)^2} {1\over18} \right)
     \intdx \ (\nabla^\mu U^T\nabla^\nu U)(\nabla_\mu U^T\nabla_\nu U)
\nonumber\\
&-& \left( 3 \lambdam + {1\over(4\pi)^2} {3\over4} \right)
     \intdx \ (\nabla^\mu U^T\nabla_\mu U)(\chi_0^T U) \ .
\end{eqnarray}

The first term on the right hand side of eq.~(\ref{Delta_2a}) is more
complicated,
because the next-to-leading singularity of the kernel $\langle x|\delta
D^{-1}\delta^T|y\rangle$ is relevant in this case as well. The computation of
the corresponding
contribution, which may be
based on either the short distance expansion or the calculation of
\mbox{two-,} three- and
four-point functions, can be found in the next two sections. Both methods
yield the following result at order $p^4$
\begin{eqnarray}
\lefteqn{ {-i\over 2} \Tr [ \delta D^{-1} \delta^T d_0^{-1} ] =
- 2 m^2 \left( \lambdam - {1\over(4\pi)^2} {1\over4} \right)
     \intdx \ \nabla_\mu U^T \nabla^\mu U} \hspace{0.5cm} \nonumber \\
&-& \left( {4\over3} \lambdam - {1\over(4\pi)^2} {5\over9} \right)
     \intdx \ (\nabla^\mu U^T\nabla_\mu U)^2 \nonumber\\
&+& \left( {4\over3} \lambdam - {1\over(4\pi)^2} {5\over9} \right)
     \intdx \ (\nabla^\mu U^T\nabla^\nu U)(\nabla_\mu U^T\nabla_\nu U)
\nonumber\\
&-& \left(\lambdam - {1\over(4\pi)^2} {1\over4} \right)
     \intdx \ (\chi_0^T U) (\nabla^\mu U^T\nabla_\mu U) \nonumber \\
&-&  {1\over(4\pi)^2} {1\over12}
     \intdx \ (\chi_0^T U)^2 \nonumber\\
&-& \left( \sixth \lambdam + {1\over(4\pi)^2} {1\over72} \right)
     \intdx \  U^T F_{\mu\nu} F^{\mu\nu} U \nonumber\\
&-& \left( \third \lambdam - {1\over(4\pi)^2} {11\over36} \right)
     \intdx \ \nabla_\mu U^T F^{\mu\nu}\nabla_\nu U \nonumber \\
&+&  {1\over(4\pi)^2} {1\over12}
     \intdx \ \chi_0^T\chi_0 \ .
\label{Delta_21_sec4}
\end{eqnarray}

The expressions in
eqs.~(\ref{linclass},\ref{Delta_1},\ref{d_Delta_2},\ref{Delta_21_sec4})
provide the complete
result for the expansion of the one-loop approximation to the generating
functional of the linear sigma model in powers of the momentum up to the order
$p^4$. Now one can read off the relations between the low energy constants and
the parameters of the linear sigma model. The contributions of order $p^2$
occurring in eqs.~(\ref{linclass},\ref{Delta_1},\ref{Delta_21_sec4}), which
have
to be matched with the effective
Lagrangian $\lag_2$ given in eq.~(\ref{eqmonlag}), can be brought into the form
\begin{equation}\label{order2contrib}
F^2 \intdx\left[ \half (\nabla_\mu U^T\nabla^\mu U) + \chi^TU\right] \ ,
\end{equation}
provided that we set
\begin{eqnarray}
F^2 & = & \left({m^2\over g}\right)\left[1 - g \left( 12 \lambdam -
{1\over(4\pi)^2} \right) \right]\\
\chi &=& \chi_0\left[1 + g \left( 6 \lambdam - {1\over(4\pi)^2} \right)
\right]. \label{chi_chi0}
\end{eqnarray}
The relation between $\chi$ and $\chi_0$ then allows one to determine the
remaining low energy constants by matching the contributions of order $p^4$
occurring in eqs.~(\ref{nonlag4}) with the effective Lagrangian $\lag_4$:
\begin{eqnarray}
d_1 &=&  {1\over4g} - {17\over3}\lambdam - {1\over(4\pi)^2}{35\over36}
\nonumber\\
d_2 &=&  {2\over3}\lambdam - {1\over(4\pi)^2} {11\over18} \nonumber\\
d_3 &=&  {1\over4g} - {9\over2}\lambdam - {1\over(4\pi)^2}{11\over6}
\nonumber\\
d_4 &=&  {1\over2g} -10\lambdam - {3\over(4\pi)^2}         \nonumber\\
d_5 &=&  {-\sixth} \lambdam - {1\over(4\pi)^2} {1\over72}\nonumber\\
d_6 &=&  {-\third} \lambdam + {1\over(4\pi)^2} {11\over36} \nonumber\\
d_7 &=&  {1\over(4\pi)^2} {1\over12} \nonumber\\
d_8 &=& h \ .
\label{d_i_bare}
\end{eqnarray}

Finally, we have to show that it was indeed consistent to neglect all
corrections which are due to the difference $U^A - \bar U^A$.
By comparing the equations of motion for $U^A$ and $\bar U^A$ given in
eqs.~(\ref{eqmolinU},\ref{eqmonlin}) one infers that $U^A - \bar U^A$ receives
contributions of order $p^2$, due to the fact that $R\neq1$, and also of
order $g$, since $\chi \neq \chi_0$. First of all we can use the relation
(\ref{chi_chi0}) to express $\chi_0$ by $\chi$ everywhere in the generating
functional $W_{\sigma}$. The replacement of $U^A$ by $\bar U^A$ can lead to
corrections of order $p^4$ only in the following contribution to the
generating functional:
\begin{eqnarray} \label{def_Sigma}
\Sigma[U,F,\chi] & \doteq  &
\int\d^dx {F^2\over 2} \left( \N_\mu U^T\N^\mu U  + 2 \chi^T U\right)
\nonumber \\
& + & \left({1\over 4g}\right)\int\d^d x
\left( \N_\mu U^T\N^\mu U  + \chi^T U\right)^2 \ .
\end{eqnarray}
In all the remaining terms one can safely replace $U^A$ by $\bar U^A$, since
the corresponding corrections to the generating functional are of the order
$g$ and $p^6$ which is beyond the accuracy of our  calculation.
In order to show that
\be \label{Sigma_U_barU}
\Sigma[U,F,\chi] = \Sigma[\bar U,F,\chi] + \order(g) + \order(p^6) \ ,
\ee
we make an intermediate step to separate the expansions in $p^2$ and in
$g$. Let $V^A$ be the stationary point of the functional
$\Sigma[V,F,\chi]$, subject to the condition $V^T V = 1$. Since $U^A$
itself extremizes the functional $\Sigma[U,{m/ \sqrt{g}},\chi_0]$, it can
easily be seen that $U^A - V^A = \order(g)$ and, therefore,
\be \label{Sigma_U_V}
\Sigma[U,F,\chi] = \Sigma[V,F,\chi] + \order(g) \ ,
\ee
since at the stationary point only corrections proportional to the square of
$U - V$ can enter.
Similarly, because $\bar U^A$ extremizes the action of the
nonlinear sigma model (cf. eq.~(\ref{eqmonlag})) we have
$\bar U^A - V^A = \order(p^2)$, from which we get
\be \label{Sigma_barU_V}
\Sigma[\bar U,F,\chi] = \Sigma[V,F,\chi] + \order(p^6) \ .
\ee
Putting eqs.~(\ref{Sigma_U_V}) and (\ref{Sigma_barU_V}) together we have
proven the claim in eq.~(\ref{Sigma_U_barU}).
The main point in the whole analysis is that, at the stationary point, the
replacement of $V$ by $V + \delta V$ can only lead to corrections which are of
second order in $\delta V$.

\rsection{Employing the Heat Kernel}

In Euclidean space the short distance properties of the differential operator
$D$,
\begin{equation}
    	D = -D_\mu D_\mu + \sigma, \qquad D_\mu = \partial_\mu + \Gamma_\mu \ ,
\end{equation}
are governed by the Laplacian $\Box$. In $d$-dimensional Euclidean space one
has
\begin{equation}
\langle x | e^{ \lambda \Box} | y \rangle =
             (4 \pi \lambda)^{-d/2} e^{- z^2 / 4 \lambda}  \ ,
\end{equation}
where $z = x-y$. In order to determine the properties of the operator $D$ at
short distances, we define~\cite{heatkernel}
\begin{equation} \label{HK}
\langle x | e^{- \lambda D} | y \rangle =  (4 \pi \lambda)^{-d/2}
\ e^{-z^2 / 4 \lambda} \ \ H(x|\lambda|y) \ .
\end{equation}
The heat kernel $H(x|\lambda|y)$ satisfies the differential equation
\begin{equation} \label{diffeq_H}
\left( {\partial \over \partial \lambda} + {1 \over \lambda} z_\mu D_\mu -
D_\mu D_\mu + \sigma \right) H(x|\lambda|y) 	 =  0
\end{equation}
with the boundary condition
\begin{equation}
				H(x|0|x) 	 =  1 \ .
\end{equation}
Inserting the Taylor expansion
\begin{equation}  \label{taylorH}
H(x|\lambda|y) = \sum_{n=0}^{\infty} \lambda^n \ H_n(x|y)
\end{equation}
into the differential equation~(\ref{diffeq_H}), one obtains the following
recursive relations for the heat coefficients $H_n$:
\begin{equation} \label{recursion_Hn}
\begin{array}{rcl}
\left( 1+n + z_\mu \stackrel{\rightarrow}{D_\mu}  \right) H_{n+1}
+ \left( - \stackrel{\rightarrow}{D_\mu}
           \stackrel{\rightarrow}{D_\mu} + \sigma \right) H_n 	& = & 0 \\
		z_\mu \stackrel{\rightarrow}{D_\mu} H_0	& = & 0.
\end{array}
\end{equation}
Note that the covariant derivative $D_\mu$ which appears on the
left hand side of the heat coefficients acts to the right on the space
coordinate $x$. In the foregoing equation we have shown this explicitly by the
arrows. In the following we will furthermore
use the covariant derivative $\stackrel{\leftarrow}{D_\mu}$,
\begin{equation}
		\stackrel{\leftarrow}{D_\mu} \doteq
               \stackrel{\leftarrow}{\partial_\mu} - \Gamma_\mu  \ ,
\end{equation}
which will always appear on the right hand side of the heat coefficients and
act to the left on the coordinate $y$. Applying the derivatives
$\stackrel{\rightarrow}{D_\mu}$ and $\stackrel{\leftarrow}{D_\mu}$ repeatedly
to the relations in eq.~(\ref{recursion_Hn}) one obtains the following results
for the heat coefficients
\begin{eqnarray}
H_0(x|x) 	& = &	1 \label{H0result}\\
H_1(x|x) 	& = & 	- \sigma \\
H_2(x|x) 	& = & 	{1 \over 12} \Gamma_{\mu \nu} \Gamma_{\mu \nu}
			+ \half \sigma^2 - {1 \over 6}
			\left[ D_\mu , \left[ D_\mu , \sigma \right] \right]
\label{H2result}\\
\left.(\stackrel{\rightarrow}{D_\mu} H_0)\right|_{x=y} &=&
\left.\left(H_0 \stackrel{\leftarrow}{D_\nu}\right)\right|_{x=y}
 =  	0 \label{DHnull} \\
\left.(\stackrel{\rightarrow}{D_\mu}
 \stackrel{\rightarrow}{D_\nu} H_0)\right|_{x=y} &=&
\left.\leftHright(_\mu,_\nu)\right|_{x=y}
 =  	\half \Gamma_{\mu \nu} \ , \label{DDHnull}
\end{eqnarray}
where
\begin{equation}
	\Gamma_{\mu\nu} \doteq \left[ D_\mu, D_\nu \right] \ .
\end{equation}
The kernel $\langle x|\delta D^{-1}\delta^T|y\rangle$ occurring in the traces
of eq.~(\ref{Delta_2a}) can be rewritten in the form
\begin{equation} \label{delta_D_deltaT2}
\langle x | \delta D^{-1} \delta^T | y \rangle =
\left( (D_\mu f_\mu)^T +  2 f_\mu^T \stackrel{\rightarrow}{D_\mu} \right)_x
\langle x | D^{-1} | y \rangle
\left( 2 \stackrel{\leftarrow}{D_\nu} f_\nu + (D_\nu f_\nu) \right)_y
\end{equation}
which in turn can be expressed in terms of the heat kernel, due to the relation
\begin{equation} \label{inversD}
\langle x | D^{-1} | y \rangle =  \int_0^{\infty} {\d \lambda \over
(4 \pi \lambda)^{d/2} } \ e^{- z^2 / 4 \lambda} \ \ H(x|\lambda|y) \ .
\end{equation}
It is convenient to define yet another kernel $h(x|\lambda|y)$ by
\begin{equation} \label{delta_D_deltaT}
\langle x | \delta D^{-1} \delta^T | y \rangle \doteq \int_0^{\infty}
{\d \lambda \over (4 \pi \lambda)^{d/2} } \ e^{-z^2 / 4 \lambda} \ \
h(x|\lambda|y) \ .
\end{equation}
If we use the following representation for the massive propagator,
\begin{equation}
G_M(z) = \int_0^{\infty} {\d \rho \over (4 \pi \rho)^{d/2}}
\ e^{{-z^2\over4 \rho} - 2 m^2 \rho} \ ,
\end{equation}
the first term on the right hand side of eq.~(\ref{Delta_2a}) can be brought to
the form
\begin{equation} \label{Delta_21a}
 \Tr [\delta D^{-1} \delta^T d_0^{-1}]  = {1 \over (4 \pi)^d}
\int_0^{\infty} {\d \lambda \d \rho \over \lambda^{d/2} \rho^{d/2}} \ \int
\d^d x \d^d y
\ e^{-{z^2 \over 4} \left( {1\over \lambda} + {1\over \rho} \right) - 2 m^2
\rho}
\ h(x|\lambda|y) \ .
\end{equation}

The kernel $h(x|\lambda|y)$ is readily determined from
eqs.~(\ref{HK},\ref{delta_D_deltaT2},\ref{inversD},\ref{delta_D_deltaT}) and
turns out to be
\begin{eqnarray}
\lefteqn{h(x|\lambda|y)  =
	2 f_\mu^T \left(\stackrel{\rightarrow}{D_\mu} H\right) (D_\nu f_\nu)
     +  2 (D_\mu f_\mu)^T \left(H \stackrel{\leftarrow}{D_\nu}\right)f_\nu}
\nonumber \\
&&{} +
	4 f_\mu^T \left(\stackrel{\rightarrow}{D_\mu} H
                    \stackrel{\leftarrow}{D_\nu} \right)  f_\nu
	+ (D_\mu f_\mu)^T H (D_\nu f_\nu)
	+ {2\over \lambda} f^T_\mu H f_\mu \nonumber \\
&&{}-	{z_\mu \over \lambda} \Biggl\{
	2 f_\mu^T \left(H \stackrel{\leftarrow}{D_\nu}\right)f_\nu
	- 2 f_\nu^T \left(\stackrel{\rightarrow}{D_\nu} H\right) f_\mu
	+ f_\mu^T H (D_\nu f_\nu)\nonumber \\
&&{}	- (D_\nu f_\nu)^T H f_\mu
	 \Biggr\}
        - {z_\mu z_\nu \over \lambda^2} f_\mu^T H f_\nu \ . \label{hxly}
\end{eqnarray}
Note that in this equation all factors appearing on the right hand side of
the heat kernel depend on the space coordinate $y$ while all factors appearing
on the left hand side depend on $x$.
The exponential factor in the integrand in eq.~(\ref{Delta_21a}) cuts off
contributions
from large values of $z$. Since the kernel $h(x|\lambda|y)$ is a smooth
function for slowly varying external fields, it admits a Taylor expansion in
powers of $z$. The expansion of the last term in eq.~(\ref{hxly}), for example,
can be written in the form
\begin{eqnarray}
\lefteqn{f_\mu^T(x) H(x|\lambda|y) f_\nu(y) } \nonumber \\
& = & \left.\left(f^T_\mu H f_\nu\right)\right|_{x=y} +
       z_\rho\left.\left( (D_\rho f_\mu)^T H f_\nu + f_\mu^T
(\stackrel{\rightarrow}{D_\rho} H) f_\nu\right)\right|_{x=y} \nonumber \\
& &\left.\mbox{} + {z_\rho z_\sigma\over2}
        \left((D_\rho D_\sigma f_\mu)^T H f_\nu +
      2 (D_\rho f_\mu)^T (\stackrel{\rightarrow}{D_\sigma} H) f_\nu
      + f_\mu^T (\stackrel{\rightarrow}{D_\rho}
                 \stackrel{\rightarrow}{D_\sigma} H)
f_\nu\right)\right|_{x=y} \nonumber \\
& & \mbox{} + \ldots
\end{eqnarray}
If we furthermore insert the Taylor expansion~(\ref{taylorH}) for the
heat kernel and
keep only terms which are relevant at order $p^4$ we obtain
\begin{eqnarray}
\lefteqn{h(x|\lambda|y) = (D_\mu f_\mu)^T H_0 (D_\nu f_\nu) + 4 f_\mu^T
\leftHright(_\mu,_\nu) f_\nu} \nonumber\\
& &   +{2\over\lambda}f_\mu^T H_0 f_\mu + 2 f_\mu^T H_1 f_\mu
   +{z_\rho z_\sigma\over\lambda} \left[ (D_\rho D_\sigma f_\mu)^T H_0 f_\mu
                        + f_\mu^T \leftleftH(_\rho,_\sigma) f_\mu \right]
\nonumber \\
& & - {z_\mu z_\rho\over \lambda}
    \left[ (D_\rho f_\mu)^T H_0 (D_\nu f_\nu) -(D_\rho D_\nu f_\nu)^T H_0
f_\mu\right.\nonumber \\
& & \left.\qquad + 2 f_\mu^T \leftHright(_\rho,_\nu) f_\nu
              -2f_\nu^T \leftleftH(_\rho,_\nu)f_\mu \right] \nonumber \\
& & -{z_\mu z_\nu\over\lambda^2} f_\mu^T H_0 f_\nu
           -{z_\mu z_\nu\over\lambda} f_\mu^T H_1 f_\nu    \nonumber \\
& & -{z_\mu z_\nu z_\rho z_\sigma\over 2\lambda^2}
           \left[ (D_\rho D_\sigma f_\mu)^T H_0 f_\nu
                        + f_\mu^T \leftleftH(_\rho,_\sigma) f_\nu \right] \ .
\label{H1occ}
\end{eqnarray}
Note that we have already omitted all terms that are odd in $z$ or contain one
of the factors given in eq.~(\ref{DHnull}) because they will drop out anyway.
Now one can perform the Gaussian integration over $z$ and use the
relations~(\ref{H0result}) to (\ref{DDHnull}) for the heat coefficients to
obtain (up to partial integrations)
\begin{eqnarray} \label{Delta_21d}
\lefteqn{   \Tr [\delta  D^{-1} \delta^T d_0^{-1} ]   =
{1 \over (4 \pi)^{d/2}} \intdx
\int_0^{\infty} {\d \lambda \d \rho \over (\lambda + \rho)^{d/2}} \
e^{- 2 m^2 \rho}}  \\
& & \Biggl[
2 f_\mu^T \Gamma_{\mu \nu} f_\nu - 2 f_\mu^T {\sigma} f_\mu +
(D_\mu f_\mu)^T (D_\nu f_\nu)  + {2\over\lambda+\rho} f_\mu^T f_\mu
\nonumber \\
& &{} - { 2 \rho \over (\lambda + \rho)} \left\{
(D_\mu f_\nu)^T (D_\mu f_\nu) + 2 (D_\mu f_\mu)^T (D_\nu f_\nu) +
2 f_\mu^T \Gamma_{\mu \nu} f_\nu -  f_\mu^T \sigma f_\mu \right\} \nonumber \\
&&{} + {2 \rho^2 \over (\lambda + \rho)^2} \left\{ (D_\mu f_\nu)^T
			(D_\mu f_\nu) +
 (D_\nu f_\mu)^T (D_\mu f_\nu) + (D_\mu f_\mu)^T (D_\nu f_\nu) \right\}
\Biggr].  \nonumber
\end{eqnarray}
To bring this result into the form of eq.~(\ref{Delta_21_sec4}) we make use of
the relation given in eq.~(\ref{ffidentity}) as well as of the following
identity
\begin{equation} \label{x1ident}
     (D_\mu f_\nu)^T (D_\rho f_\sigma)
 = \nabla_\mu\nabla_\nu U^T \nabla_\rho \nabla_\sigma U -
  (U^T \nabla_\mu\nabla_\nu U)(U^T \nabla_\rho \nabla_\sigma U) \ .
\end{equation}
Note that some of the terms in this equation do not occur in the effective
Lagrangian given in eq.~(\ref{nonlag4}). These terms can be expressed in terms
of the complete set of independent invariants that occur in the Lagrangian. In
order to eliminate redundant terms we proceed as discussed at the end of
section 3. Using the definition of the field strength $F_{\mu\nu}$ given in
eq.~(\ref{Fdef}) one can eliminate the various dependent terms of the form
$\nabla_\mu\nabla_\nu U^T \nabla_\rho\nabla_\sigma U$ in favour of only
$\nabla_\mu\nabla_\mu U^T \nabla_\nu\nabla_\nu U$ and some other terms, that
are already present in the effective Lagrangian. Since we have furthermore used
the equations of motion of the effective Lagrangian to reduce the number of
independent terms in $\lag_4$, we should do the same in this case as well. From
the equation of motion~(\ref{eqmolinU}) of the linear sigma model we infer a
relation similar to the one given in eq.~(\ref{eqmocond2}),
\begin{equation} \label{x2ident}
\nabla_\mu\nabla_\mu U^T\nabla_\nu\nabla_\nu U - (\nabla_\mu U^T\nabla_\mu U)^2
 - \chi_0^T\chi_0 + (\chi_0^T U)^2 = 0 \ ,
\end{equation}
which holds
up to corrections of order $p^6$ which can be neglected. Thus, using the
identities~(\ref{x1ident},\ref{x2ident}) to eliminate all terms of the form
$\nabla_\mu\nabla_\nu U^T \nabla_\rho\nabla_\sigma U$ we obtain the result
given in eq.~(\ref{Delta_21_sec4}).

\rsection{Loops}

In the preceding two sections we have evaluated all one-loop contributions in
the matching relation~(\ref{match2})  which determine the low energy constants
of the effective Lagrangian at order $p^4$ by means of the short-distance
expansion in configuration space.  In this section we want to compare this
method with the evaluation by means of loop-integrals in momentum space.
We shall see that at order $p^4$ this involves  tadpole graphs and two-,
three-, and four-point functions. Since we want to present a general
comparison between
these two methods -- short-distance expansion versus loop-integrals -- we will
not discuss all contributions in detail.

First, we will discuss the evaluation of the second term on the left hand side
of eq.~(\ref{match2}) which can be expanded as shown in eq.~(\ref{lndetd}). The
second term on the right hand side of this equation is a simple tadpole graph
of the massive propagator as shown explicitly in eq.~(\ref{lndetd_2}). The
third term in eq.~(\ref{lndetd}) involves a two-point function with two massive
propagators. In momentum-space it is given by (cf. eq.~(\ref{lndetd_3}))
\be \label{twopoint}
- {i\over4} \int {d^dp_1 \over (2\pi)^d} \sigma_M(p_1) \sigma_M(-p_1)
         \int {d^dp \over (2\pi)^d} {1\over (M^2 - p^2) (M^2 - (p + p_1)^2)}
\ .
\ee
The loop-integral is readily evaluated and is regular at $p_1 = 0$. Hence, it
admits an expansion for small momenta,
\bea  \label{B_M_M}
\lefteqn{ \int {d^dp\over (2\pi)^d} {1\over (M^2 - p^2) (M^2 - (p + p_1)^2)} }
\nonumber \\
&&= -i \left( 2 \hat\lambda_0(M^2) - {1\over 16\pi^2}
\left\{ \rho \, \ln \left( {\rho -1\over \rho +1} \right) + 1 \right\} \right)
\\
&&=  -i \left( 2 \hat\lambda_0(M^2) + {1\over 16\pi^2} \right) + \order(p_1^2)
\ ,
\label{B_M_M_b}
\eea
where
\be
\rho \doteq \sqrt{1 - {4 M^2 \over p_1^2}} \ .
\ee
At order $p^4$ only the leading term of the expansion of this two-point
function is relevant.
Thus one obtains the same result as given in eq.~(\ref{lndetd_3}).
This may be compared with the short-distance expansion used in
eq.~(\ref{lndetd_3}), where the integrand
$\sigma_M(x) \sigma_M(x+y)$ is expanded in $y$. In both cases only the leading
term is relevant at order $p^4$.

To discuss the evaluation of the contributions from the third term on the left
hand side of eq.~(\ref{match2}), it is convenient to introduce the
following short-hand notation for the $n$-point functions. We define the
general two-point function by
\be
B_{\mu_1\ldots\mu_n}(p_1;m_1,m_2) \doteq {1\over i} \int {d^dp\over (2\pi)^d}
{p_{\mu_1} \ldots p_{\mu_n} \over (m_1^2 - p^2) (m_2^2 - (p + p_1)^2)} \ .
\ee
The function $B(p_1;m_1,m_2)$ denotes the integral where no momenta
appear in the numerator, e.g. $B(p_1;M,M)$ is given explicitly in
eq.~(\ref{B_M_M}). Similarly we denote the general three-point function by
$C_{\mu_1\ldots\mu_n}(p_1,p_2;m_1,m_2,m_3)$  and the general four-point
function by $D_{\mu_1\ldots\mu_n}(p_1,p_2,p_3;m_1,m_2,m_3,m_4)$.

Whereas the kernel $\langle x| d^{-1}|y\rangle $ is related to the propagator
$G_M$ of the massive particle, the kernel $\langle x| \Dh^{-1} |y \rangle $,
which enters in eq.~(\ref{Delta_2a}), leads to propagators  $G_0$ of the
massless particles in the loops.
We can formally rewrite the operator $D$ in the form (cf. eq.~(\ref{def_D}))
\be \label{D_formal}
\Dh  \doteq  \Box \ (1 + \Box^{-1} ((D_\mu\Gamma^\mu)
                       + 2 \Gamma_\mu \partial^\mu + \sigma)) \ .
\ee
Therefore we get
\bea
\lefteqn{ \hspace*{-4mm} \langle x | ( \Dh^{-1} )^{ik} | y \rangle =
\delta^{ik} G_0(x-y) }
\label{inversD_2}     \\
& - &  \int\d^dz G_0(x-z) \left( (D_\mu\Gamma^\mu)^{ik}(z)
		+ 2 \Gamma_\mu^{ik}(z) \partial^\mu + \sigma^{ik}(z)
\right) G_0(z-y) \nonumber \\
& + & 4 \int\d^dz\d^du \, \Gamma_\mu^{ij}(z) \Gamma_\nu^{jk}(u) \,
G_0(x-z) \partial^\mu G_0(z-u) \partial^\nu G_0(u-y)
+ \ldots \ .\nonumber
\eea

In the last two traces on the right hand side of eq.~(\ref{Delta_2a}) only the
first term in expansion~(\ref{inversD_2}) is relevant. Note that this
term also determines the leading singularity of the kernel $\langle x|
\delta \Dh^{-1} \delta^T |y \rangle $. The relevant contribution to the
second term on the right hand side of eq.~(\ref{Delta_2a}) is given by the
integral (cf.~eq.~(\ref{Delta_2a_2}))
\begin{equation} \label{secondterm}
-2 \int {d^dp_1 \over (2\pi)^d} {d^dp_2 \over (2\pi)^d} f^{\mu T}(p_1+p_2)
f^\nu(-p_1) \sigma_M(-p_2) C_{\mu\nu}(p_1,p_2;0,M,M) \ ,
\end{equation}
which involves two massive and one massless propagator.
The loop-integral which contributes to the third term in eq.~(\ref{Delta_2a})
involves two massive and two massless propagators and is of the form
(cf.~eq.~(\ref{Delta_2a_3}))
\begin{eqnarray} \label{thirdterm}
\lefteqn{\hspace{-1.5cm} 4 \int {d^dp_1 \over (2\pi)^d}
	{d^dp_2 \over (2\pi)^d} {d^dp_3 \over (2\pi)^d}
   f^{\mu T}(p_1+p_2+p_3) f^\nu(-p_1) f^{\rho T}(-p_2) f^\sigma(-p_3)
} \nonumber\\
& & \qquad\qquad  D_{\mu\nu\rho\sigma}(p_1,p_2,p_3;0,M,0,M) \ .
\end{eqnarray}
In expressions~(\ref{secondterm}) and~(\ref{thirdterm}) the external fields
account already for four powers in the momenta so that only the leading terms
in the expansions of the function  $C_{\mu\nu}(p_1,p_2;0,M,M)$ and
$D_{\mu\nu\rho\sigma}(p_1,p_2,p_3;0,M,0,M)$, which are of order~1, are
relevant.
Using the expressions for the leading terms as given in the appendix, one
obtains the same results as given in eq.~(\ref{d_Delta_2}).

The evaluation of the first term on the right hand side of eq.~(\ref{Delta_2a})
by means of the short distance expansion was more complicated, because in this
case the next-to-leading singularity of the kernel  $\langle x| \delta \Dh^{-1}
\delta^T |y \rangle $  is relevant as well. In the language of loop-integrals
this translates to the fact, that all three terms in
expansion~(\ref{inversD_2}) of the kernel  $\langle x| \Dh^{-1} |y \rangle $
need to be taken into account. Hence, at order $p^4$ the first term on the
right hand side of eq.~(\ref{Delta_2a}) receives
contributions from two-, three-, and four-point functions. In particular, one
obtains
\begin{eqnarray}\label{fourt2}
\lefteqn{\hspace{-0.2cm}
\half \int{d^dp_1\over(2\pi)^d} \left\{ b^T(p_1) a(-p_1) B(p_1;0,M)
 +  2i b^T(p_1) f_\mu (-p_1) B^\mu(p_1;0,M) \right. } \\
&  & \qquad\mbox{}-\left. 2i f_\mu^T (p_1) a(-p_1) B^\mu(p_1;0,M)
 + 4  f_\mu^T(p_1) f_\nu(-p_1) B^{\mu\nu}(p_1;0,M)\right\} \ ,\nonumber
\end{eqnarray}
\begin{eqnarray}
\lefteqn{\hspace{-0.8cm}-2\int{d^dp_1\over(2\pi)^d}{d^dp_2\over(2\pi)^d}
 \Bigg\{\Big[ f_\mu^T(p_1+p_2) \left((D_\rho\Gamma^\rho)(-p_1) +
\sigma(-p_1)\right) f_\nu(-p_2)} \nonumber \\
& & \qquad\qquad\mbox{} + b^T(p_1+p_2) \Gamma_\mu(-p_1) f_\nu(-p_2)
\nonumber\\
& & \qquad\qquad\mbox{} - f_\mu^T(p_1+p_2)\Gamma_\nu(-p_1) a(-p_2)\Big]
C^{\mu\nu}(p_1,p_2;0,0,M)  \nonumber\\
& & \qquad\mbox{} - 2i f_\mu^T(p_1+p_2) \Gamma_\nu(-p_1) f_\rho(-p_2)
	\Big[ C^{\mu\nu\rho}(p_1,p_2;0,0,M) \nonumber\\
& & \qquad\qquad\mbox{}		+ p_1^\nu C^{\mu\rho}(p_1,p_2;0,0,M)
		+ p_1^\rho C^{\mu\nu}(p_1,p_2;0,0,M)\Big]\Bigg\},
\label{fourt3}
\end{eqnarray}
and
\begin{eqnarray}\lefteqn{\hspace{-2.6cm}
-8\int{d^dp_1\over(2\pi)^d}{d^dp_2\over(2\pi)^d}{d^dp_3\over(2\pi)^d}
 f_\mu^T(p_1+p_2+p_3) \Gamma_\nu(-p_1) \Gamma_\rho(-p_2)
f_\sigma(-p_3)}\nonumber\\
& & \qquad\qquad\qquad D^{\mu\nu\rho\sigma}(p_1,p_2,p_3;0,0,0,M) \
,\label{fourt4}
\end{eqnarray}
where
\begin{equation}
 a =  (D_\mu f^\mu) - 2 \Gamma_\mu f^\mu \ , \qquad
 b^T  =  (D_\mu f^\mu)^T + 2 f_\mu^T \Gamma^\mu \ .
\end{equation}
The leading singularity of the first term on the right hand side of
eq.~(\ref{Delta_2a}) is determined by the d'Alembertian, i.e.
\begin{equation}
- 2 i \Tr [ f^\mu\stackrel{\rightarrow}{\partial_\mu}
 {1\over\Box} \stackrel{\leftarrow}{\partial_\nu} f^\nu  d_0^{-1} ] \ .
\end{equation}
In expression~(\ref{fourt2}) it is described by the term involving the
two-point
function $B_{\mu\nu}(p_1;0,M)$. Accordingly, the first two terms of the
low-energy expansion of this two-point function are relevant at order $p^4$. In
this expansion the term of order 1 describes the contribution from the leading
singularity while the term of order $p^2$ contributes to the next-to-leading
one. The contributions from the heat coefficient $H_1$, occurring in
eqs.~(\ref{H1occ}), are readily identified as well. They are given in
expression~(\ref{fourt3}) by the term involving the quantity $\sigma$.

In this way one recovers the same result as was obtained by using the
heat-kernel method. However, the evaluation of the term
$- {i\over 2} \Tr (\delta \Dh^{-1} \delta^T d_0^{-1})$ by means of
loop-integrals is painstaking, since due to the formal inversion of the
operator
$\Dh$ in
eq.~(\ref{inversD_2}) the covariant form of the  final result is not obvious.

Finally, we should mention that all three- and four-point functions with two or
more massless propagators have infrared singularities in 4 dimensions. They
would have to be taken care of, if we were to extract the contributions of
order $p^6$ from the loop integrals, using the next terms in the low energy
expansion of these functions.  For instance, the next-to-leading order
contribution of  $C_{\mu\nu}(p_1,p_2;0,0,M)$ would enter at this order.  The
analysis shown in the appendix reveals that the low momentum  expansion of this
function is of the form
\be \label{logocc}
C_{\mu\nu}(p_1,p_2;0,0,M) = P_{\mu\nu}(p_1,p_2) + Q_{\mu\nu}(p_1,p_2) \,
\ln({-p_1^2\over M^2}) + \mbox{higher order,}
\ee
where $P_{\mu\nu}$ and $Q_{\mu\nu}$ are polynomials of second degree in $p_1$
and $p_2$. Only $P_{\mu\nu}$ contains a term of order 1, which leads to a
contribution of order $p^4$ through eq.~(\ref{fourt3}). However, terms
involving logarithms, like the one in eq.~(\ref{logocc}), do not contribute to
the effective Lagrangian.

Note that we already encountered such terms in our calculation. They
were hidden in the nonlocal determinant $\ln\det \Dh$, which cancelled in
the comparison  of the generating functionals of the linear sigma model and of
the effective theory in eq.~(\ref{match1}).
A similar situation would occur if we were to determine the contributions at
order $p^6$. Note that in this case one has to calculate the
generating functional in the linear sigma model as well as in the effective
theory up to the two-loop level. Therefore additional terms would occur beyond
$\ln\det \tilde D$, which represents the one loop graphs only. In this way
one gets nonlocal terms in the linear sigma model and in the effective theory
from loops, where a Goldstone boson propagates over long distances, leading to
the above mentioned infrared problems. These nonlocal terms would, however,
cancel in the matching of the generating functionals and the remaining local
terms would lead to the coefficients $d_i$ at order $p^4$ and to analogous
coefficients at order $p^6$.

The short distance expansion is only appropriate to get
these local terms. If we use, for instance, a representation of the
propagator $\langle x | \Dh^{-1} | y \rangle$ by the heat-kernel as in
eq.~(\ref{inversD}), it only makes
sense to expand the heat-kernel for small values of the parameter $\lambda$,
which corresponds to small distances $x - y$. Infrared singularities on the
other hand show up at large $\lambda$ where the expansion does not make sense,
especially if massless particles are present.

\rsection{Renormalization}

In order to obtain finite results for physical quantities one has to
renormalize the bare constants $g, m$ and $h$ before the regulator can be
removed (there is no wave-function renormalization at the order considered
here, such that the external field $f^A$ does not require renormalization).
This will indeed render the linear $O(N)$ sigma model finite,
because it is a renormalizable theory. To analyze the ultraviolet behaviour of
the determinant of the differential operator $\tilde D$ given in
eq.~(\ref{diffoplin}) we again employ the heat kernel method. In Euclidean
space-time the $d$-dimensional determinant may be defined as
\begin{eqnarray} \label{heatdet}
\ln\det\tilde D & = & - \int_0^\infty{\d\lambda\over\lambda} \Tr
e^{-\lambda\tilde D}   \\
   & = & -(4\pi)^{-d/2}\int_0^\infty \d \lambda \lambda^{-1-d/2}
             \int\d^dx \ \tr \tilde H(x|\lambda|x) \ ,
\end{eqnarray}
where the $\tilde H$ is the heat kernel of the operator $\tilde D$ defined in
eq.~(\ref{HK}). In this representation the divergences which are related to the
ultraviolet behaviour show up at the lower end of the integration over
$\lambda$. For $d=0,2,4,\ldots$ the determinant has poles. In order to
identify the residues of these poles, we split the integration over $\lambda$
into an integral from $0$ to $\lambda_0$ and a remainder. Using the Taylor
series expansion~(\ref{taylorH}) we obtain
\begin{eqnarray}
   \half\ln\det\tilde D & = &{1\over d}\int \d^dx \ \tr \tilde H_0(x|x)
                  + {1\over d-2}{1\over4\pi}\int\d^dx \ \tr \tilde H_1(x|x)
\nonumber \\
                  & & \mbox{} + {1\over d-4}{1\over16\pi^2}
                                           \int\d^dx \ \tr \tilde H_2(x|x)
			+ \ldots
\end{eqnarray}
With the result given in eq.~(\ref{H2result}), the the pole-term at
$d=4$ turns out to be given by
\begin{equation}
 {1\over d-4}{1\over16\pi^2} \int\d^dx \left\{
   {1\over12} \tr \left( F_{\mu\nu}F_{\mu\nu} \right)
    + \half \tr ( \tilde\sigma^2 ) \right\} \ .
\end{equation}
Note that the differential operator $\tilde D$ is positive in the $O(N)$
symmetric phase of the linear sigma model. In this case the integrand in
eq.~(\ref{heatdet}) is exponentially damped as $\lambda$ goes to infinity,
which expresses the fact that there are no infrared divergences in this
phase. Since the divergence structure of a renormalizable field
theory is the same in both, the symmetric and the spontaneously broken phase,
the pole is removed by the following renormalization prescriptions:
\begin{eqnarray}
         g & = & g_r \left( 1 - 2 (N+8) g_r(\lambdam+\delta g)\right)
+\order(g_r^3) \label{g_r} \\
         m^2 & = & m_r^2 \left( 1 - 2 (N+2) g_r(\lambdam+\delta m^2)\right) +
\order(g_r^2) \\
          h & = & h_r + {1\over 12} \lambdam + \order(g_r) \ . \label{h_r}
\end{eqnarray}
These relations introduce the five finite but otherwise arbitrary constants
$m_r, g_r, \delta m^2, \delta g$ and $h_r$.  In the case of a (massive) Yang
Mills theory, where the gauge field $F_\mu$ is a dynamical field, $h$
represents the inverse square of the gauge coupling constant. In the present
context, where $F_\mu$ is an external field, loops involving gauge fields do
not occur. The renormalization of $h$ encountered here only accounts for the
renormalization of the gauge coupling constant  which are due to loops of the
field $\phi^A$.
The finite renormalization constants $\delta m^2$ and $\delta g$ are in general
eliminated by requiring two renormalization conditions. In addition to that, we
will also replace the remaining two independent parameters $m_r$ and $g_r$ by
two other quantities, the physical mass $M$ of the heavy particle and the low
energy constant $F$, whose physical meaning will be discussed at the end of
this section.

For the rest of this section we will use Minkowski space notation.
The mass $M$ is determined by the pole position of the connected
two-point function
\begin{equation}
\left.-i{\delta^2\over\delta f^0(x)\delta f^0(y)}
W_\sigma\left[F_\mu,f]\right]\right|_{F_\mu=f=0}
=     \left<0|T(\phi^0(x)\phi^0(y))|0\right>_{conn}  \ .
\end{equation}
Note that we have embedded the unbroken symmetry group $O(N-1)$ in such a way
that it does not act on the first component $\phi^0$ of the $N$-component field
$\phi^A$. Thus, $\phi^0$ describes the massive mode while the remaining $N-1$
components $\phi^i$, $i=1,\ldots,N-1$, represent the Goldstone bosons.
In order to evaluate the pole position, we use eqs.~(\ref{linfunc1})
and~(\ref{tildeexpansion}) as well as the solution $R$ of the equations of
motion~(\ref{eqmolinR}),
\begin{eqnarray}
\lefteqn{R(x)-1 = \int\d^dy G_M(x-y) \chi_0^0(y) } \\
&- & 3 m^2\int\d^dy\d^dv\d^dw G_M(x-y) G_M(y-v)G_M(y-w) \chi_0^0(v) \chi_0^0(w)
+ \ldots \ .
\nonumber
\end{eqnarray}
One obtains the following result for the pole position:
\bea
  M^2 - i M\Gamma & = & 2 m_r^2\Biggl[ 1 - 2(N+2)g_r \delta m^2
- {1\over16\pi^2} g_r
\Biggl\{ 9\left(1-{\pi\over\sqrt{3}}\right) \nonumber \\
&& \qquad \mbox{}+   (N-1)(1+i\pi)\Biggr\}\Biggr] \ ,
\eea
where $M$ and $\Gamma$ are the mass and width of the heavy particle.
In terms of the renormalized quantities $m_r$ and $g_r$, the low-energy
constant
$F$ turns out to be
\begin{equation}
   F^2 = \left({m_r^2\over g_r}\right)\left[1 + 2 g_r({1\over32\pi^2} -
(N+2) \delta m^2 + (N+8)\delta g) \right] \ .
\end{equation}
Thus, one can eliminate the two renormalization constants $\delta m^2$ and
$\delta g$ and replace the two parameters $m_r$ and $g_r$ by the physical mass
$M$ and the low-energy constant $F$ by requiring
\begin{eqnarray}
      M^2     & = & 2 m_r^2 \ , \nonumber \\
      F^2     & = & {m_r^2\over g_r} \label{fixingM_F} \ .
\end{eqnarray}
Thus, $\delta m^2$ and $\delta g$ are given by
\begin{eqnarray}
  (N+2) \delta m^2 & = & - {1\over 32
\pi^2}\left(9\left(1-{\pi\over\sqrt{3}}\right)+N-1\right) \ , \\
  (N+8) \delta g & = &  (N+2)\delta m^2 - {1\over32\pi^2} \ .
\end{eqnarray}
The low-energy constants as well as the field $\chi$ turn out to be
\begin{eqnarray}
\chi&=& {1\over F}\left(1- {M^2\over F^2}{1\over 64\pi^2} \right) f \nonumber
\\
d_1 &=&  {3N-10\over6}\lambda_0  + {F^2\over2M^2} - {1\over192\pi^2}\left(
		  {116+9N-27\sqrt{3}\pi\over3}
		- (3N-10) \ \ln{M^2\over\mu^2} \right) \nonumber\\
d_2 &=&  {2\over3}\lambda_0  - {1\over96\pi^2}\left( {11\over3}
		- 2 \ \ln{M^2\over\mu^2} \right) \nonumber\\
d_3 &=&  {N-1\over2}\lambda_0  + {F^2\over2M^2} - {1\over64\pi^2}\left(
		  {49+3N-9\sqrt{3}\pi\over3}
		- (N-1) \ \ln{M^2\over\mu^2} \right) \nonumber\\
d_4 &=&  (N-2)\lambda_0 + {F^2\over M^2} - {1\over32\pi^2}\left(
		  15+N-3\sqrt{3}\pi
		- (N-2) \ \ln{M^2\over\mu^2} \right)  \nonumber\\
d_5 &=&  -{1\over6}\lambda_0 - {1\over 192\pi^2} \left( {1\over6}
		+ \ \ln{M^2\over\mu^2} \right)   \nonumber\\
d_6 &=&  -{1\over3}\lambda_0 + {1\over 96\pi^2} \left({11\over6}
		-  \ \ln{M^2\over\mu^2} \right) \nonumber\\
d_7 &=&  {1\over192\pi^2}\nonumber\\
d_8 &=&  {1\over 12} \lambda_0 + h_r +
		{1\over 384\pi^2} \ \ln{M^2\over\mu^2} \ .
\label{d_i_ren}
\end{eqnarray}
Note that the low-energy constants are of the form
\begin{equation}
	d_i =  \delta_i \lambda_0 + \bar d_i \ ,
\end{equation}
where the $\bar d_i$ are finite and expressed in terms of physical quantities.
It is furthermore important to note that the matching relation~(\ref{Zequal})
between the linear sigma model and the effective theory determines both
contributions to the low-energy constants, the finite parts $\bar d_i$ as well
as the poles in $d=4$. These pole terms, $\delta_i \lambda_0$, occur in
tree-diagrams of order $p^4$ and exactly cancel the corresponding poles in the
contributions from loops with vertices from $\lag_2$, as represented by the
determinant in eq.~(\ref{genfuncp4}). Since the effective theory describes the
low-energy structure of the linear sigma model, the renormalization of the
latter must indeed render the former finite as well.

Let us now compare the results for the low-energy constants given in
eqs.~(\ref{d_i_bare}) and~(\ref{d_i_ren}) with those of
refs.~\cite{GasserLeutChpt,Espriu,HRM}\footnote{
The low-energy constants $d_i$ are related to those in
ref.~\cite{GasserLeutChpt} by
$l_3 = d_3 - d_4, h_1 = d_4 + d_7, h_2 = d_8,$ and $l_i = d_i$ for $
i=1,2,4,5,6,$ and to those in ref.~\cite{Espriu} by $4 L_1 = d_1, 4 L_2 = d_2,
- 2 L_9 = d_6,$ and $L_{10} = d_5.$}.
Our results
in eq.~(\ref{d_i_bare}), where the low-energy constants are expressed in
terms of bare quantities, agree with those in eq.~(B.11) of
ref.~\cite{GasserLeutChpt}. However, the renormalization condition used in that
article differs from the one given in eq.~(\ref{fixingM_F}) above, which has to
be taken into account if the results
of our eq.~(\ref{d_i_ren}) are compared with eq.~(B.12) of
ref.~\cite{GasserLeutChpt}.

The authors of ref.~\cite{Espriu} integrate only over the heavy scalar field in
the linear sigma model and, thus, obtain the functional $\Gamma_{eff}$ as
defined in eq.~(\ref{def_Gamma}) above. The expansion of this functional in
terms of local quantities defines another set of parameters, $d_i^\Gamma$,
which can be inferred from eqs.~(\ref{linclass}) and~(\ref{Delta_1}). We
obtain, for example, $d_2^\Gamma = d_5^\Gamma = d_6^\Gamma = 0$, which agrees
with the results of ref.~\cite{Espriu}. The constants~$d_2,d_5$ and~$d_6$, on
the other hand, receive contributions from mixed loops and are different
from zero. The effective Lagrangian, however, which describes the
low-energy structure of the linear sigma
model involves the low-energy constants $d_i$ as given in
eqs.~(\ref{d_i_bare}) and~(\ref{d_i_ren}) and not the parameters
$d_i^\Gamma$.

The analysis of the low-energy constants in ref.~\cite{HRM} includes the
effects of dynamical gauge fields as well. In that article, the contributions
from mixed loops are taken into account and, furthermore, it was found that
loops containing gauge fields do not contribute to the low-energy constants at
order $p^4$. For a more detailed comparison between the low-energy constants
obtained in refs.~\cite{GasserLeutChpt} and \cite{HRM} the reader is referred
to ref.~\cite{HRM}.

Let us conclude this section with two remarks about the low-energy constant
$F$. At the order we are considering here it is identical with the constant
$F_\pi$ which describes the coupling of the conserved currents to
the Goldstone bosons. Recall that the gauge fields $F_\mu^\alpha$ couple to
the conserved currents $J_\mu^\alpha$ given in eq.~(\ref{conservedcurrents}).
By choosing a suitable basis for the $N-1$ Goldstone boson states
$\left.|\pi^k\right>$ Lorentz invariance implies that these currents have the
following matrix elements
\begin{equation}  \label{Fpidef}
\left<0|J_\mu^\alpha(0)|\pi^k(p)\right> = i \delta^{\alpha k} F_\pi p_\mu \ ,
\end{equation}
which defines the coupling constant $F_\pi$. Note that only the $N-1$ currents
which correspond to broken generators of the $O(N)$ symmetry couple to the
Goldstone bosons.
According to definition (\ref{Fpidef}), the residue of the Goldstone boson
pole of the two-point function
\begin{equation}
\left.-i{\delta^2\over\delta
F^\alpha_\mu(x)\delta F^\beta_\nu(y)}
W_\sigma\left[F_\mu,f\right]\right|_{F_\mu=f=0}
=     \left<0|T(J^\alpha_\mu(x)J^\beta_\nu(y))|0\right>_{conn}
+\mbox{contact terms}
 \ ,
\end{equation}
is given by $F_\pi^2$. Note
that the contact terms do not contribute to the pole position.
Since $F_\pi$ is a low-energy constant, it may as well be determined from our
representation of the generating functional $W_{eff}$.
The explicit calculation yields
\begin{equation}
     F = F_\pi \ .
\end{equation}

At tree-level, the coupling constant $F_\pi$ is identical with the
vacuum expectation value $v$ of the field $\phi^0$, which is given by
\begin{equation}
v  \doteq  \left<0|\phi^0(x)|0\right>_{conn} = \left.{\delta\over\delta
f^0(x)} W_\sigma\left[F_\mu,f\right]\right|_{F_\mu=f=0} \ .
\end{equation}
Beyond tree-level, however, both quantities differ by corrections of order $g$
and the relation is given by
\begin{equation}
 v  =  F\left(1 - {M^2\over F^2}{1\over64\pi^2}\right)  \ .
\end{equation}
This reflects the fact that the constant $F_\pi$ is not renormalized, because
the normalization of the currents is fixed by the commutation relations. The
vacuum expectation value $v$, on the other hand, depends on the normalization
of the fields and, hence, on a convention.

Finally we want to make a comment on the validity of our first-order
perturbative calculation. The physics of the linear $O(N)$ sigma model in the
spontaneously broken phase can adequately be described by an effective theory,
which contains only the Goldstone bosons, if energy and momenta are small
compared to the mass of the heavy particle, i.e.~if $p^2/M^2 << 1$. The
perturbative expansion in the coupling constant $g$, on the other hand, breaks
down if this mass is too large, since $g \sim M^2$ according to
eq.~(\ref{fixingM_F}).  As an example we consider the Standard Model values
$N = 4$ and  $v = 246~GeV$. In this case perturbation theory in the parameter
$g$ can no longer be applied if the mass is of the order of 1 TeV.

\rsection{Summary}

In this work we have analyzed the low-energy structure of the linear $O(N)$
sigma model from an effective field theory point of view. In the spontaneously
broken phase the spectrum of this theory consists of one massive mode and $N-1$
Goldstone bosons. At low energies, physical phenomena are dominated by the
light particles of a theory, while the presence of heavier species manifests
itself rather indirectly. In the low energy region, the nontrivial structure of
the linear sigma model is related to the singularities of Green's functions
which are associated with the Goldstone bosons. The effects of the heavy
particle, on the other hand, admit a Taylor expansion and show up as simple
power-like contributions. Hence, the low energy structure of this theory can
adequately be described by an effective theory which contains only the
Goldstone bosons. In the Lagrangian which describes the
effective theory, effects of the heavy particle manifest themselves as an
infinite tower of nonrenormalizable interactions which are accompanied by an
infinite series of dimensionful low energy (coupling) constants. The
requirement that both, the effective and the full theory describe the same low
energy physics in principle determines the relationships between all
these low energy constants and the parameters of the underlying theory. In more
mathematical terms, the equivalence between the effective theory and the full
theory requires that corresponding Green's functions have the
same low energy structure. This can be expressed in a compact way as
\begin{equation} \label{Wequal2}
	W_{\sigma}\left[F_\mu,f\right] = W_{eff}\left[F_\mu,\chi\right]  \ ,
\end{equation}
where  $W_{\sigma}[F_\mu,f]$ and $W_{eff}[F_\mu,\chi]$ are the generating
functionals
of the linear sigma model and of the effective theory, respectively.
They depend on a set of external vector fields $F_\mu$ and scalar fields
$\chi$ or $f$, which are coupled to the matter fields. Derivatives of these
functionals with respect to the vector fields $F_\mu$ generate Green's
function of the $O(N)$ currents, while derivatives with respect to the scalar
fields $\chi$ or $f$ generate Green's functions of the matter fields.
Equation~(\ref{Wequal2}) should not be understood as an identity but rather as
an
asymptotic equality in the low energy region. It determines the low energy
constants as well as the relation between the external scalar fields $\chi$
and $f$. More explicitly, eq.~(\ref{Wequal2}) requires the following relation
between full path integrals
\begin{equation}\label{con1}
        \int \d\mu[U] e^{i\int\d^d \lag_{eff}(U,F_\mu,\chi)} =
	\int \d\mu[U]\d\mu[R] e^{i\int\d^d \lag_{\sigma}(R,U,F_\mu,f)} \ ,
\end{equation}
where the effective Lagrangian $\lag_{eff}(U,F_\mu,\chi)$ depends on the
external sources and on the fields
$U$ of the Goldstone bosons, while the Lagrangian of the linear sigma model
$\lag_{\sigma}(R,U,F_\mu,f)$ depends on the field
$R$ of the heavy degree of freedom as well. At the classical level, this
relation between integrals is equivalent to the following relation between
integrands
\begin{equation}\label{con2}
        e^{i\int\d^d \lag_{eff}(U,F_\mu,\chi)} =
	\int \d\mu[R] e^{i\int\d^d \lag_{\sigma}(R,U,F_\mu,f)} \ .
\end{equation}
This equivalence, however, is lost if quantum corrections are taken into
account. In other words, beyond tree-level, the correct matching condition is
given by relation~(\ref{con1}) between the full integrals. In this sense,
integrating out heavy particles should rather be understood as replacing the
full theory by an effective one which describes the same low energy physics.

At low energies, Green's functions receive nonlocal contributions which are
related to the propagation of light particles over large distances. Since both,
the effective and the full theory have the same light particle content, all
nonlocal contributions drop out of the matching relation. The remaining local
terms involve the propagation of heavy particles over short distances. It has
been shown in this work how the heat kernel technique can be used to evaluate
the short-distance expansion of these terms and thus determine the functional
relationships between the low energy constants and the parameters of the full
theory. These relationships can, however, also be evaluated by means of loop
integrals in momentum space. In order to present a detailed description of both
methods we have explicitly evaluated the effective Lagrangian for the linear
$O(N)$ sigma model in the spontaneously broken phase up to the order $p^4$. In
contrast to the short distance expansion, the evaluation based on
loop-integrals is not covariant during intermediate steps of the calculation.
The momentum space picture is useful, however, to understand why
eq.~(\ref{con2}) does not define the correct effective Lagrangian beyond
tree-level. The local terms in the matching condition receive quantum
corrections from two types of Feynman diagrams, cf.~eq.~(\ref{match2}), those
which involve propagators only of the heavy degrees of freedom and those which
involve propagators of both, heavy and light particle species. The latter ones
are obviously missed in eq.~(\ref{con2}). Finally we have expressed the low
energy constants in terms of the mass $M$ of the heavy particle and the decay
constant $F$.

\section*{Acknowledgements}

We are grateful to H.~Leutwyler for many enlightening discussions and a
critical
reading of this manuscript. One of us (A.S.) is furthermore indebted to
M.~Dugan, H.~Georgi and S.~Vandermark for interesting discussions. He also
wants to thank the members of the Harvard Physics Department for their kind
hospitality.

\section*{Appendix}
\renewcommand{\theequation}{A.\arabic{equation}}
\setcounter{equation}{0}

In this appendix we briefly want to discuss how the low energy expansion of
$n$-point functions can be constructed. As a particular example we shall
consider the two three-point functions $C_{\mu\nu}(p_1,p_2;0,M,M)$ and
$C_{\mu\nu}(p_1,p_2;0,0,M)$. Note, that the latter one has infrared
singularities while the former one is regular for small momenta. In Euclidean
space-time one has
\begin{eqnarray}
\lefteqn{C_{\mu\nu}(p_1,p_2;0,m_2,m_3)} \\
& = &   \int {\d^d p\over (2\pi)^d} {p_\mu p_\nu \over
                p^2( (p+p_1)^2 + m_2^2) ((p+p_1+p_2)^2 + m_3^2) } \ .
\nonumber
\end{eqnarray}
A convenient Feynman parametrization is given by
\begin{eqnarray}
\alpha\left[(1-\beta)\left( (p+p_1)^2 + m_2^2\right) + \beta\left(
(p+p_1+p_2)^2 + m_3^2\right)\right] &&\\
 + (1-\alpha)p^2
&\doteq&p^2 + 2 p k + \Delta \ , \nonumber
\end{eqnarray}
with
\begin{eqnarray}
k & = & \alpha(p_1 + \beta p_2) \\
\Delta & = & \alpha\left[(1-\beta) \left(m_2^2 + p_1^2\right) +
\beta\left(m_3^2 +  (p_1+p_2)^2\right)\right] \ .
\end{eqnarray}
The integration over $p$ yields
\begin{equation}
C_{\mu\nu} =
\int_0^1 \d\alpha \alpha \int_0^1\d\beta \left(
	k_\mu k_\nu F(0) + \half \delta_{\mu\nu} F(1) \right) \ ,
\end{equation}
where
\begin{equation}
     F(l) = {1\over (4\pi)^{d/2}}{\Gamma(3-l-{d\over2})\over
(\Delta-k^2)^{3-l-d/2} } \ .
\end{equation}
In the following we will furthermore use dimensionless momenta
\begin{equation}
q_i  \doteq {p_i\over M} \ .
\end{equation}

\noindent
In the case of $C_{\mu\nu}(p_1,p_2;0,M,M)$, $m_2=m_3=M$ and we have
\begin{equation}
\Delta - k^2 = \alpha M^2\left[ 1 + f_1(\alpha,\beta,q_1,q_2) \right] \ ,
\end{equation}
where the function
\begin{equation}
f_1(\alpha,\beta,q_1,q_2) \doteq (1-\beta)q_1^2 + \beta (q_1+q_2)^2 - \alpha
(q_1+\beta q_2)^2
\end{equation}
is of order $q^2$. Thus, the three-point function
$C_{\mu\nu}(p_1,p_2;0,M,M)$ is regular for small momenta and the low energy
expansion is obtained by expanding the integrand in powers of $f_1$.
At leading order we get
\begin{eqnarray}
C_{\mu\nu}(p_1,p_2;0,M,M) &=& -{1\over2}\delta_{\mu\nu} \left\{\lambdaM +
		{1\over32\pi^2}{1\over2} \right\}.
\end{eqnarray}

\noindent
In the case of $C_{\mu\nu}(p_1,p_2;0,0,M)$, $m_2=0$ \& $m_3=M$ and we get
\begin{equation}
\Delta - k^2 = \alpha M^2 F(\alpha,\beta,q_1,q_2) ,
\end{equation}
where
\begin{equation}
F(\alpha,\beta,q_1,q_2)\doteq \beta + (1-\beta)q_1^2 + \beta (q_1+q_2)^2
	 	- \alpha (q_1+\beta q_2)^2 \ .
\end{equation}
Since $F(\alpha,\beta,q_1,q_2)$ vanishes iff $\beta=0$ and $\alpha=1$, we
encounter an endpoint singularity in this case. Therefore  the three-point
function $C_{\mu\nu}(p_1,p_2;0,0,M)$ will have a branch point at $q_1^2 = 0$.
In order to analyze the behaviour of this function at small momenta, we
separate the term linear in $\beta$ and $1-\alpha$ as follows
\begin{equation}
F(\alpha,\beta,q_1,q_2) = \left( \beta(1+q_2^2)  + (1-\alpha) q_1^2 \right)
(1 + f_2(\alpha,\beta,q_1,q_2)) \ ,
\end{equation}
where the quantity
\begin{equation}\label{f2def}
f_2(\alpha,\beta, q_1, q_2) \doteq {\beta\over \beta(1+q_2^2)
                   + (1-\alpha) q_1^2}
\left( -\beta\alpha q_2^2 + 2 (1-\alpha)  q_1 q_2\right)
\end{equation}
is of order $q^2$ in the region of integration. Note, that this
function is well-behaved even in the limit $\beta\rightarrow 0^+$ and
$\alpha\rightarrow 1^-$.
Thus we can expand this integrand in powers of $f_2$. In this way one obtains
\begin{eqnarray}\lefteqn{
\int_0^1 \d\alpha \alpha \int_0^1\d\beta k^\mu k^\nu F(0)}\\
& = & {1\over16\pi^2}\left[ q_1^\mu q_1^\nu \left( {11\over18} - {1\over3}
\ln{q_1^2} \right) + {1\over6} q_2^\mu q_2^\nu + {1\over3} (q_1^\mu q_2^\nu +
q_1^\nu q_2^\mu) \right] + \ldots \nonumber
\end{eqnarray}
and
\begin{eqnarray}\lefteqn{
 \int_0^1 \d\alpha \alpha \int_0^1\d\beta	 F(1)}\\
& = & -\lambdaM + {1\over32\pi^2} \left[
\half - {11\over18} q_1^2 - {2\over 3} q_2^2 - {2\over3} q_1 q_2
+ {1\over3} q_1^2 \ln{q_1^2} \right] + \ldots \ .\nonumber
\end{eqnarray}

For completeness sake we finally present all the terms in the low
energy expansions of the $n$-point functions that we use in section 7
(in Minkowski metric):
\begin{eqnarray}
B(p_1;M,M) &=& -2 \lambdaM - {1\over 16\pi^2}   \\
B(p_1;0,M) &=& -2 \lambdaM  \\
B^\mu(p_1;0,M) &=& p_1^\mu \left( \lambdaM +
		{1\over 32\pi^2} {1\over2} \right) \\
B^{\mu\nu}(p_1;0,M) &=&
- {M^2 \over 2} g^{\mu\nu} \left( \lambdaM -
		{1\over 32\pi^2} {1\over 2} \right) \nonumber \\
& & +  {p_1^2 \over 6} g^{\mu\nu} \left( \lambdaM +
		{1\over 32\pi^2} {1\over 6} \right) \nonumber \\
& & - {2\over 3} p_1^\mu p_1^\nu \left( \lambdaM +
		{1\over 32\pi^2} {2\over 3} \right) \\
C^{\mu\nu}(p_1,p_2;0,M,M) &=& {1\over2}g^{\mu\nu} \left(\lambdaM +
		{1\over32\pi^2} {1\over2}\right) \\
C^{\mu\nu}(p_1,p_2;0,0,M) &=& {1\over2}g^{\mu\nu} \left(\lambdaM -
		{1\over32\pi^2}{1\over2}\right) \\
C^{\mu\nu\rho}(p_1,p_2;0,0,M) &=& -{g^{\mu\nu} p_1^\rho\over3} \left(
	\lambdaM - {1\over 32 \pi^2}{1\over3} \right) \nonumber \\
 && -  {g^{\mu\nu} p_2^\rho\over6} \left(
	\lambdaM + {1\over 32 \pi^2}{1\over6} \right) \nonumber \\
& &  - {\rm permutations\ of\ }(\mu\nu\rho) \\
D^{\mu\nu\rho\sigma}(p_1,p_2,p_3;0,M,0,M)
& = & -{1\over 12} \left(g^{\mu\nu} g^{\rho\sigma} + g^{\mu\rho} g^{\nu\sigma}
+
g^{\mu\sigma} g^{\nu\rho} \right) \nonumber \\
& & \qquad \left( \lambdaM + {1\over 32\pi^2}{1\over6} \right)   \\
D^{\mu\nu\rho\sigma}(p_1,p_2,p_3;0,0,0,M)
&= & -{1\over 12} \left(g^{\mu\nu} g^{\rho\sigma} + g^{\mu\rho} g^{\nu\sigma} +
g^{\mu\sigma} g^{\nu\rho} \right)\nonumber\\
& & \qquad \left( \lambdaM - {1\over 32\pi^2}{5\over6} \right) \ .
\end{eqnarray}

\end{document}